\def\COMBINEDVERSION{1}

\documentclass[pdflatex,sn-mathphys-num,oneside]{sn-jnl}

\usepackage{graphicx}%
\usepackage{multirow}%
\usepackage{amsmath,amssymb,amsfonts}%
\usepackage{amsthm}%
\usepackage{mathrsfs}%
\usepackage[title]{appendix}%
\usepackage{xcolor}%
\usepackage{textcomp}%
\usepackage{manyfoot}%
\usepackage{booktabs}%
\usepackage{algorithm}%
\usepackage{algorithmicx}%
\usepackage{algpseudocode}%
\usepackage{listings}%
\usepackage{soul}
\usepackage{needspace}
\usepackage{longtable}
\usepackage{array}
\makeatletter
\@ifundefined{@kernel@refstepcounter}{}{%
  \patchcmd{\caption}{\refstepcounter}{\@kernel@refstepcounter}{}%
    {\PackageError{dwc}{Caption hyperlink patch failed}{Check hyperref version.}}}
\makeatother

\ifdefined\COMBINEDVERSION
    \usepackage{bibunits}
    \makeatletter
    \def\@bibunitname{\jobname-bu\the\@bibunitauxcnt}
    \newcommand{\bibliographynamespace}[1]{%
      \def\@extra@binfo{#1}%
      \def\@extra@b@citeb{#1}}
    \makeatother
\else
    \usepackage{xr-hyper}
\fi

\theoremstyle{thmstyleone}%

\theoremstyle{thmstyletwo}%

\theoremstyle{thmstylethree}%
\begin{document}


\title[Article Title]{Rethinking electrified transportation: The case for dynamic wireless charging}

\author[1]{\fnm{Dipanjan} \sur{Ghose}}\email{dghose@purdue.edu}

\author[1]{\fnm{Junjie} \sur{Qin}}\email{jq@purdue.edu}

\author*[2]{\fnm{S.} \sur{Sivaranjani}}\email{sseetha@purdue.edu}

\affil[1]{
\orgdiv{Elmore Family School of Electrical and Computer Engineering},
\orgname{Purdue University},
\orgaddress{\state{IN}, \country{USA}}}

\affil*[2]{
\orgdiv{Edwardson School of Industrial Engineering},
\orgname{Purdue University},
\orgaddress{\state{IN}, \country{USA}}}

\abstract{
Long-term infrastructure investment often requires choosing between an established technology and a capital-intensive alternative that gains value as demand matures. For electric vehicle (EV) charging, this trade-off is between incumbent stationary charging technology and emerging dynamic wireless charging (DWC). DWC could fundamentally alter transportation electrification by enabling high-throughput charging in motion and reducing reliance on large onboard batteries; however, its high upfront cost makes the timing and location of deployment critical. Overcommitting to stationary charging stations (SCS) risks technological lock-in and forgoing the benefits of DWC, whereas premature DWC deployment can strand capital in underutilized corridors. Through a joint investment assessment framework coupling DWC, SCS, and supporting grid infrastructure across nearly 49,000 miles of the US Interstate freight network, we find  that higher utilization and longer project lifetimes rapidly shift investment toward DWC, whose share rises sharply from 48\% at a  project lifetime of 30 years to 100\% at 32 years. Contiguous DWC segments can share grid connections, enabling them to provide over six times the rated charging capacity of SCS while requiring 46\% less  investment in new distribution-grid infrastructure. We also find that a combined DWC and SCS portfolio reaches a maximum lifetime value of USD 60 billion, 63\% above the case with only SCS. We further identify specific traffic demand, construction, and grid access conditions that shape where each technology becomes advantageous, yielding actionable thresholds for SCS and DWC deployment that directly inform long-term EV charging infrastructure investment and policy. 
}


\maketitle


\ifdefined\COMBINEDVERSION
    \bibliographynamespace{main}
    \begin{bibunit}[sn-mathphys-num]
\fi
The rapid electrification of road transportation is increasing the scale and urgency of charging needs, making today’s deployment decisions consequential for meeting future demand. Global electric vehicle (EV) sales exceeded 20 million in 2025, comprising a quarter of all new cars sold worldwide, and the number of EVs per public charging point is projected to rise by nearly 75\% by 2035 \cite{IEA2026GlobalEVOutlook}. Although scaling stationary charging station (SCS) networks has been the conventional response to rising charging demand, dynamic wireless charging (DWC) is emerging as a fundamentally different charging paradigm \cite{Elkholy25,Panchal2018WirelessChargingReview}. DWC can simultaneously deliver energy to multiple vehicles while they travel on electrified roadways, offering high charging throughput and reducing downtime \cite{ghose2025,Kim2025DynamicWirelessCharging}. This extends the driving range, paving the way for lighter onboard batteries and more affordable EVs \cite{Liu2024BatteryCapacity}. Several DWC pilot projects are now underway worldwide, including the Arena of the Future project in Italy, in-route wireless charging for buses in Trondheim, Norway, and the Purdue-INDOT DWC roadway in West Lafayette, Indiana \cite{ElectreonItaly,ElectreonNorway,INDOTDWPT}.

SCS benefits from established financing pathways and lower upfront costs, whereas DWC is more capital intensive but can become financially viable as high throughput offsets its higher costs \cite{Haddad2022FinancialViability,majhi2022optimal,Li2023EconomicViability}. This creates a core investment trade-off, as continued reliance on SCS can lock charging networks into infrastructure sized for today's demand, while premature DWC deployment can strand capital in low-utilization corridors. At the same time, DWC adoption reduces demand at stationary chargers, while supplementary SCS lowers the marginal value of electrifying nearby roadway segments, raising the question of whether, when, and where DWC or SCS should be preferentially deployed, and how the two technologies should be optimally combined as demand matures.

As DWC and SCS serve the same underlying demand, answering this requires that the two technologies be studied as a common investment portfolio. Such a formulation should also take into account the grid constraints associated with each technology, since DWC concentrates high-capacity loads that can be supplied through contiguous segments, whereas SCS creates smaller, discrete loads that each require individual grid connections. While DWC \cite{majhi2022optimal,elmeligy2024optimal,liu2017robust} and SCS \cite{pradhan2021planning, shabbiruddin2021evsc, Janji17112021, abdullah2026geographic} have each been extensively studied in isolation, the literature on their joint planning remains limited, with existing approaches focused on localized systems or stylized formulations that are difficult to scale \cite{csonka2021optimization,xia2023market,chen2017deployment,sun_integrated_2020}. Here, we develop a joint investment model for DWC, SCS, and supporting grid infrastructure that maximizes lifetime portfolio value while capturing corridor connectivity and competition for shared charging demand, demand growth over the project lifetime, construction costs,  and grid constraints and upgrade costs.  


Our results across 48,890 miles of the US Interstate freight network  show that project lifetime and sustained EV adoption growth are primary determinants of charging portfolio composition. The transition is sharp as project lifetime increases, with DWC rising from 3.5\% of the portfolio at a 26-year lifetime to 48\% at 30 years and 100\% by 32 years, as demand growth increases its long-term value. We also test how the portfolio responds to the share of EV traffic allocated to DWC, capturing, at a high level, how future technology uptake and charging preferences could affect infrastructure value. DWC and SCS create greater value when deployed together, with the combined DWC-SCS portfolio attaining USD 60 billion in lifetime value, 63\% above the SCS-only case. Grid access further shapes the relative value of the two technologies. As contiguous electrified DWC segments can share grid connections, DWC requires 46\% less new distribution line infrastructure, as measured by line mileage, while providing nearly 1 GW of rated charging capacity, more than six times that of SCS. 

These outcomes vary substantially across the network, as traffic demand, construction costs, and grid access determine where each technology becomes advantageous. We translate these spatial deployment patterns into actionable thresholds that identify the conditions favoring DWC or SCS and how these preferences shift as deployment conditions evolve. Comparing these thresholds across locations further reveals broader deployment archetypes that favor DWC or SCS, providing transferable insights for charging infrastructure planning across roadway networks beyond the US Interstate freight network. Together, these results provide a basis for rethinking roadway electrification, demonstrating how DWC can reshape the long-term planning and deployment of future EV charging infrastructure.

\begin{figure}
    \centering
    \includegraphics[width=1\linewidth]{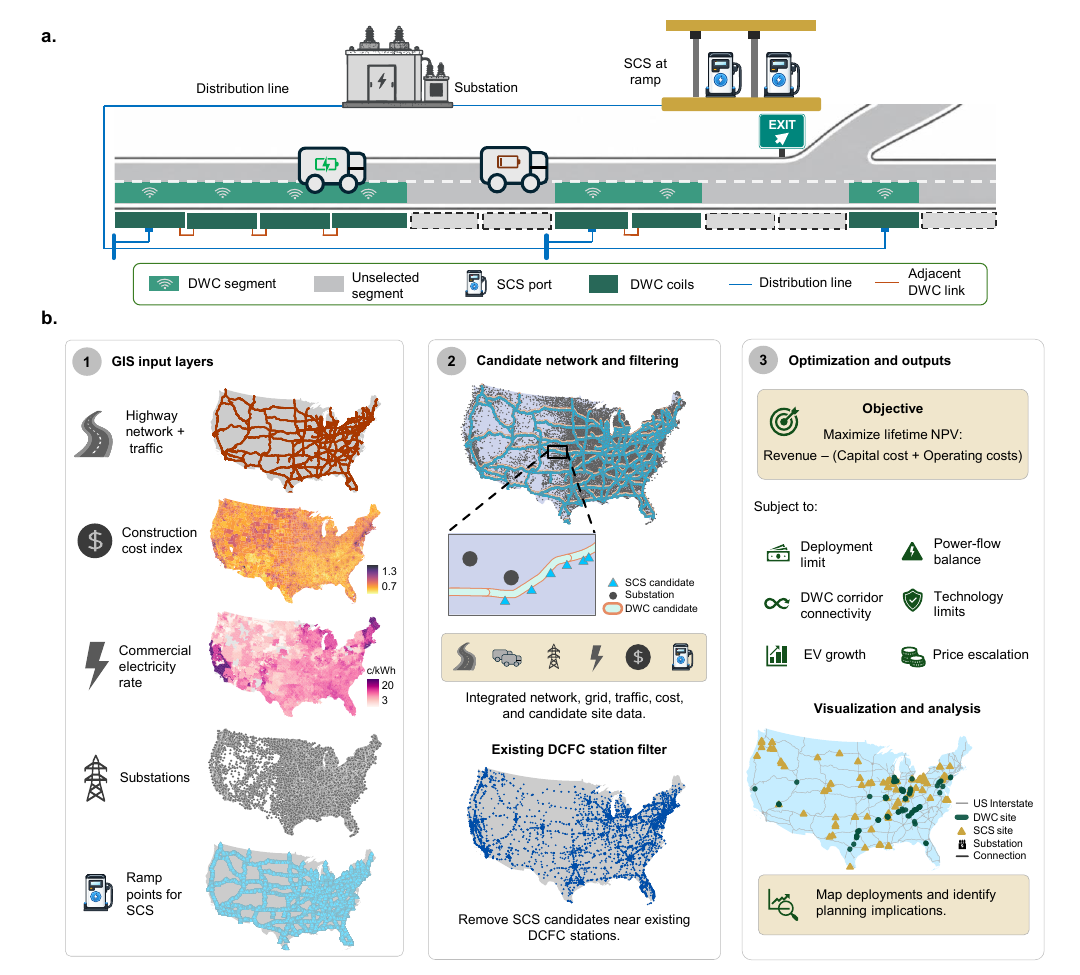}
    \caption{\textbf{Joint planning framework for DWC and SCS.}
\textbf{a.} Roadway corridors are divided into uniform candidate segments for DWC deployment, while ramp exits define candidate SCS locations serving demand from nearby segments. Contiguous DWC segments can share access to a substation through adjacent electrified segments (termed adjacent links), whereas selected SCS sites require site-specific connections to nearby substations.
\textbf{b.} The GIS-based workflow integrates highway geometry, traffic demand, construction costs, electricity prices, substation locations and candidate SCS sites. Candidate locations are filtered using existing DCFC infrastructure and passed to an optimization model that selects DWC, SCS or no deployment while  respecting all constraints.}
    \label{fig:framework}
\end{figure}
\subsubsection*{Joint investment framework for DWC and SCS infrastructure}\label{sec:Joint planning of EV charging and grid infrastructure}

We evaluate joint DWC and SCS investment across 48,890 miles of the US Interstate freight network, spanning diverse traffic, construction, and grid conditions \cite{WorldPopulationReview2026RoadNetwork,BTSMovingGoodsUS}. We discretize the network into eligible five-mile roadway segments as candidate DWC sites and use highway ramps as feasible locations for SCS, yielding 9,053 candidate DWC segments \cite{fha2016} and 17,400 candidate SCS sites. We then overlay the network with more than 40,000 substation locations \cite{xu2020ustestsystem} to represent grid access (Fig.~\ref{fig:framework}a).  Traffic demand, electricity cost, construction conditions, grid accessibility, and existing charging infrastructure are captured using annual average daily traffic (AADT), commercial electricity rates (CER), construction-cost indices (CCI), distance to the nearest substation, and nearby DC fast-charging (DCFC) stations, respectively (see Methods: \nameref{sec:GIS pipeline}). Fig.~\ref{fig:framework}b summarizes these geospatial inputs and the resulting  network.

We then maximize portfolio net present value (NPV) over the project lifetime, accounting for growing charging demand, construction and operating costs, and competition between DWC and SCS for shared EV traffic. Each deployed DWC segment and SCS site must connect to a substation, either through a dedicated distribution line or, for DWC, through adjacent electrified roadway segments. Contiguous DWC segments can therefore share grid access through adjacent links, reducing the need for direct connections, whereas each selected SCS requires a site-specific grid connection. This construction places the two charging technologies within a common lifetime investment portfolio while explicitly representing their distinct spatial and grid requirements. We represent future DWC uptake and charging preferences through exogenous DWC-SCS traffic shares, which we vary parametrically across scenarios together with project lifetime, EV-adoption growth, and technology-specific assumptions. This allows us to examine  how the relative value and spatial deployment of DWC and SCS evolve across future charging scenarios.

\begin{figure}
   \centering
   \includegraphics[width=0.75\linewidth]{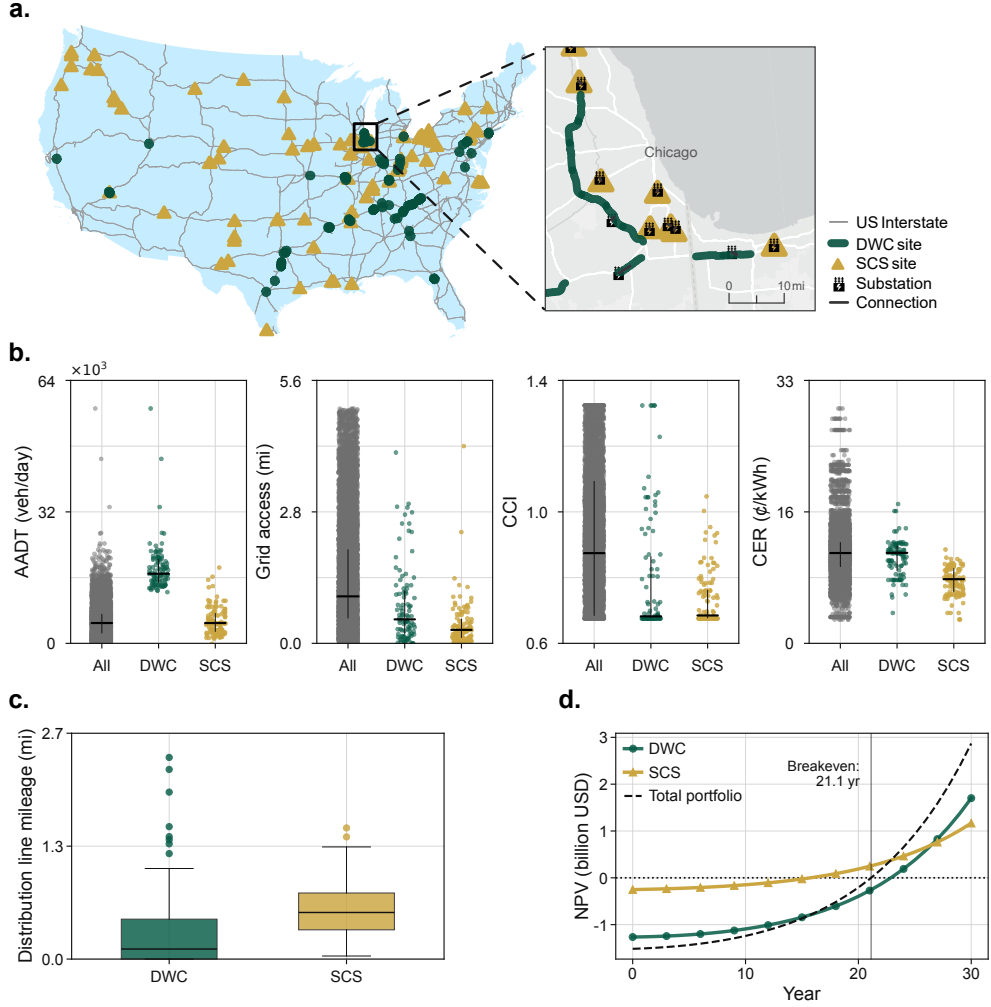}

   \caption{\textbf{Optimized DWC-SCS portfolio across the US Interstate freight network.}
   \textbf{a.} Optimal deployment showing selected DWC segments, SCS sites, nearby substations and grid connections. The baseline assumes 0.5\% initial EV penetration of freight AADT, equal shares of EV traffic assigned to DWC and SCS, a 15\% annual EV-adoption growth rate, 1.1\% annual AADT growth, a 30-year project lifetime, a deployment cap of 200 units, and a charging price of USD~0.54\,kWh\(^{-1}\) for both technologies. The Chicago inset highlights contiguous DWC corridor clusters, nearby SCS sites and their grid connections.
   \textbf{b.} Distributions of freight AADT, grid access, CCI and CER across all candidate locations and selected DWC and SCS sites.
\textbf{c.} Distribution of new grid connections for selected DWC and SCS deployments, where length denotes the new distribution line distance (mileage) required to connect the charging asset to a substation. 
   \textbf{d.} Lifetime economic performance of the optimized portfolio, showing cumulative NPV of DWC, SCS and the combined portfolio over the 30-year project lifetime, with portfolio breakeven at 21.1 years.
   Basemap sources for the Chicago inset: Esri, TomTom, Garmin, FAO, NOAA, USGS, copyright by OpenStreetMap contributors and the GIS User Community.}

   \label{fig:US_results}
\end{figure}

\subsubsection*{DWC and SCS play complementary roles in the charging portfolio}
\label{sec:Mapping optimized charging portfolios across the US}

Under the baseline scenario (see Methods: \nameref{sec:Framework Overview}), the optimized 30-year portfolio selects 96 DWC segments and 104 SCS locations, producing a near-even split between the two technologies (Fig.~\ref{fig:US_results}a). However, the two technologies contribute very differently to the resulting charging network. DWC electrifies 478.9 roadway miles and provides 957.6 MW of rated charging capacity, compared with 156.0 MW across SCS sites. The spatial characteristics of selected sites help explain why the optimized portfolio remains nearly evenly split despite DWC’s higher upfront cost (Fig.~\ref{fig:US_results}b). The selected DWC segments have a median AADT of 17,000 vehicles/day, compared with 4,970 across all candidates, and a lower median CCI of 0.70 versus 0.88 network-wide. DWC therefore concentrates in locations where higher utilization and favorable construction conditions can offset its greater capital cost. 

Grid access remains an important siting constraint, with selected locations generally closer to substations than the full candidate network. For DWC, however, contiguous electrified segments can share grid connections, allowing deployment to extend beyond locations with direct substation access.  Of the 96 selected DWC segments, 64 connect directly to substations while the remaining 32 are supplied through adjacent links. This reduces direct distribution line construction to 32.75 miles, compared with 61.11 miles for SCS, a 46\% reduction (Fig.~\ref{fig:US_results}c). 
The Chicago inset in Fig.~\ref{fig:US_results}a illustrates how this connectivity structure enables contiguous DWC deployment alongside discrete SCS sites within the same corridor.

Taken together, the contrasting utilization, capacity, and grid-connectivity characteristics of DWC and SCS translate into distinct lifetime economics (Fig.~\ref{fig:US_results}d). SCS reaches breakeven in about 16 years, compared with 23 years for DWC, reflecting its lower upfront capital requirement. DWC, however, benefits from higher utilization and charging throughput over longer horizons. The combined portfolio reaches breakeven at 21.1 years, illustrating how joint deployment balances the earlier capital recovery of SCS with the higher long-term capacity and  connectivity advantages of DWC.

\subsubsection*{Demand growth and longer project lifetimes drive a sharp transition toward DWC}
\label{sec:Demand growth and project lifetime drive the transition from SCS to DWC}

\begin{figure}
    \centering
    \includegraphics[width=1\linewidth]{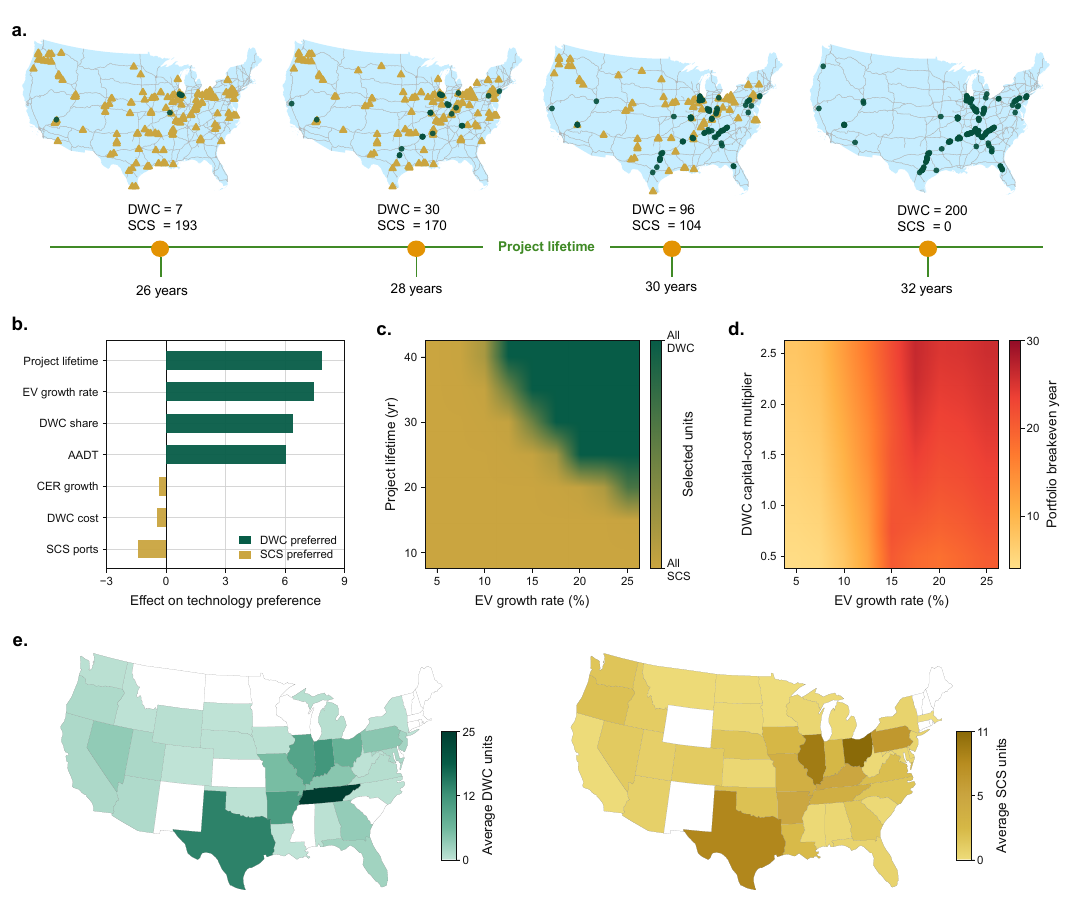}
    \caption{\textbf{Project lifetime and EV growth govern the transition between stationary and dynamic charging.}
EV growth denotes the annual growth in the EV share of roadway traffic, modeled separately from overall traffic growth.
\textbf{a.} Optimized national deployment patterns under a fixed 200-unit cap as project lifetime increases from 26 to 32 years.
\textbf{b.} Selected standardized binary logistic-regression coefficients estimated from deployed DWC and SCS observations across 1,000 GSSA runs. Positive coefficients indicate conditions favoring DWC selection and negative coefficients favor SCS.
\textbf{c.} Portfolio composition across combinations of EV growth rate and project lifetime, showing the transition from SCS- to DWC-dominated deployment.
\textbf{d.} Portfolio breakeven year across EV growth rates and DWC capital-cost multipliers. Higher DWC capital costs generally delay breakeven; localized increases near the portfolio transition reflect discrete changes in DWC-SCS composition (see Supplementary Note~\ref{sec:si-Additional Results}).
\textbf{e.} State-level numbers of DWC and SCS deployments, averaged across the planning scenarios, showing that the national transition remains spatially uneven.}
    \label{fig:sensitivityanalysis}
\end{figure}

The transition from SCS to DWC is highly sensitive to the time horizon over which infrastructure value is assessed. As project lifetime $T$ increases from 26 to 32 years, the DWC share of the optimized portfolio rises from 3.5\% to 100\%, with the sharpest transition occurring between 30 and 32 years, when its share rises from 48\% to 100\% (Fig.~\ref{fig:sensitivityanalysis}a). Longer project lifetimes therefore allow DWC's higher charging throughput to generate value over a sufficiently long horizon to offset its greater upfront cost.

We next examine whether this transition persists under simultaneous variation in demand, technology, and economic assumptions using a global scenario sensitivity analysis (GSSA) across 1,000 optimization runs (see Methods: \nameref{sec:Sensitivity Analysis}). The standardized logistic-regression coefficients in Fig.~\ref{fig:sensitivityanalysis}b identify project lifetime and EV-adoption growth as the strongest drivers favoring DWC, followed by exogenous DWC traffic share and segment AADT. Increasing the number of SCS charging ports produces the strongest opposing effect by increasing stationary-charging capacity, while higher DWC capital costs also shift deployment toward SCS. We also find that CER growth has comparatively little influence. Across the scenarios considered, the transition is therefore governed primarily by the growth and duration of charging demand rather than an increase in  electricity prices.

The interaction between demand growth and project lifetime further sharpens this result (Fig.~\ref{fig:sensitivityanalysis}c). At approximately $T\!=\!32$ years, DWC overtakes SCS near 15\% annual EV-adoption growth, while longer project lifetimes require progressively lower EV growth for the same transition. Higher EV growth increases future charging demand, while longer lifetimes provide more time for the resulting revenues to recover DWC's higher capital cost. This transition remains visible as DWC capital costs increase, although higher costs generally delay portfolio breakeven (Fig.~\ref{fig:sensitivityanalysis}d).

The transition is also spatially uneven (Fig.~\ref{fig:sensitivityanalysis}e). Higher-utilization states shift more strongly toward DWC as demand and project lifetime increase, whereas SCS remains more prevalent in lower-utilization regions. State-level differences in traffic, grid access, and regional costs underlying these spatial patterns are summarized in Supplementary Note~\ref{sec:si-Additional Results}. This mixed regime is particularly evident in Texas, where both technologies coexist at intermediate planning horizons before DWC expands rapidly as project lifetime increases, providing a useful setting to examine the spatial and economic mechanisms underlying the transition.

\subsubsection*{High throughput and corridor connectivity drive DWC expansion}

\begin{figure}
\centering
\includegraphics[width=\linewidth]{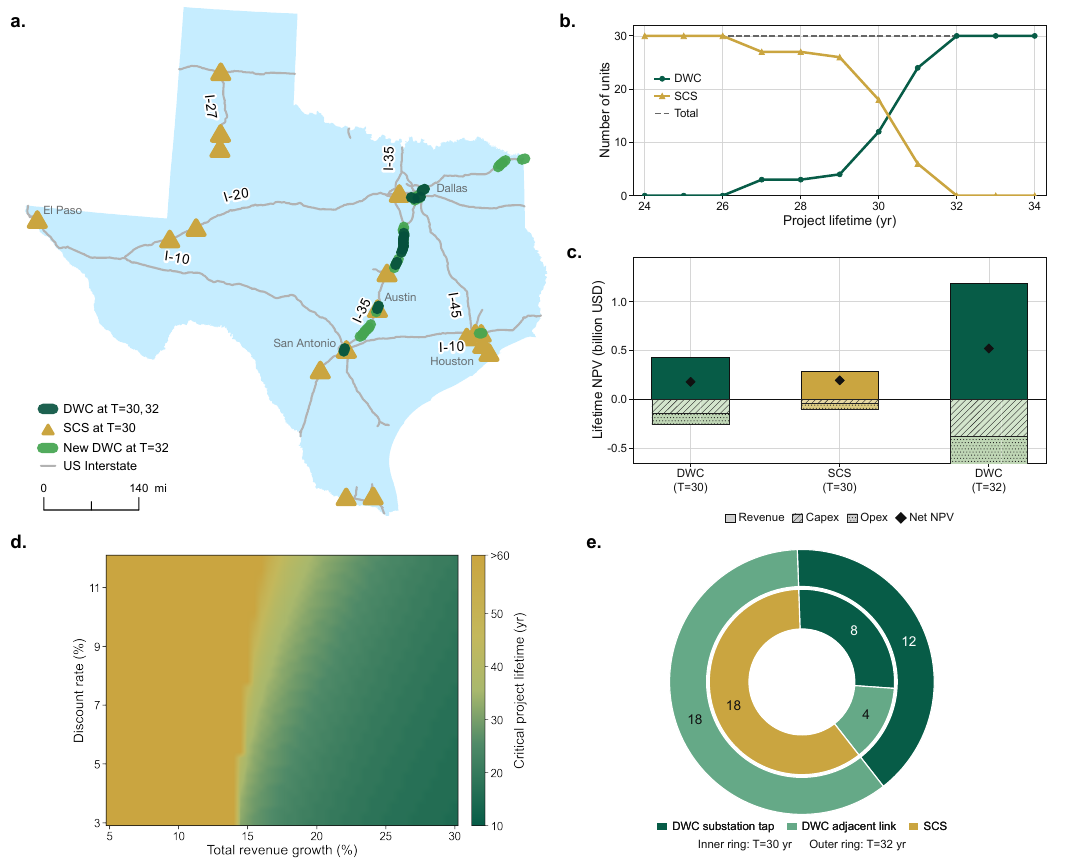}
\caption{\textbf{High throughput and corridor connectivity drive DWC expansion.} 
\textbf{a.} Optimized Texas portfolios at $T\!=\!30$ and $T\!=\!32$ years, with equal DWC and SCS EV traffic shares. At $T\!=\!32$, the original DWC segments are retained and additional contiguous segments are selected along major Interstate corridors, while no SCS sites remain. 
\textbf{b.} Numbers of DWC and SCS units selected as project lifetime increases, revealing a sharp portfolio transition between $T\!=\!30$ and $T\!=\!32$.
\textbf{c.} Lifetime revenue, capital expenditure, operating expenditure, and net NPV for DWC and SCS at $T\!=\!30$, and for DWC at $T\!=\!32$.
\textbf{d.} Critical project lifetime across total revenue-growth and discount-rate assumptions, defined as the earliest project lifetime at which representative per-unit DWC NPV exceeds representative per-unit SCS NPV. Stronger revenue growth advances this crossover, whereas discount-rate variation has a smaller effect over the tested range.
\textbf{e.} Composition of the Texas portfolio at $T\!=\!30$ and $T\!=\!32$, distinguishing SCS sites, grid-connected DWC segments, and DWC segments supplied through adjacent links. Map source for panel a: TxDOT Right of Way - Commercial Signs Regulatory Program, Texas Department of Transportation, Transportation Planning and Programming Division, Data Management.}
\label{fig:texas_transition}
\end{figure}

The preceding results show that longer project lifetimes and stronger demand growth drive a sharp transition of the charging portfolio toward DWC. To understand the economic and spatial mechanisms underlying this transition, we examine Texas, where both DWC and SCS remain well represented at intermediate planning horizons (Fig.~\ref{fig:sensitivityanalysis}e). Under a shared 30-unit deployment cap, the optimized Texas portfolio contains 12 DWC segments and 18 SCS sites at $T\!=\!30$ years. Increasing the project lifetime to $T\!=\!32$ completely eliminates SCS from the optimized portfolio while retaining the original 12 DWC segments and adding 18 new DWC segments, predominantly along contiguous Interstate corridors (Fig.~\ref{fig:texas_transition}a-b). The transition, therefore, involves not only substitution of DWC for SCS but also spatial expansion around existing electrified roadway segments. Two features of DWC help explain why this expansion becomes increasingly favorable as the project horizon lengthens.

First, the higher charging throughput of DWC increases its lifetime value as demand grows. The economic basis of this transition is evident from the NPV decomposition in Fig.~\ref{fig:texas_transition}c. At $T\!=\!30$, DWC and SCS contribute similar net NPVs of USD~0.180 billion and USD~0.195 billion, respectively, yielding a combined portfolio NPV of USD~0.375 billion. At $T\!=\!32$, the all-DWC portfolio reaches USD~0.521 billion, 39\% above the $T\!=\!30$ portfolio. As traffic and EV adoption grow, DWC's higher charging throughput allows it to serve increasing demand and generate greater value in later project years, eventually offsetting its higher upfront capital cost. Consistent with this mechanism, stronger revenue growth advances the DWC crossover to shorter project lifetimes, whereas variation in the discount rate has a smaller effect over the range considered (Fig.~\ref{fig:texas_transition}d).

Second, corridor connectivity lowers the grid-infrastructure requirements of DWC expansion. As contiguous DWC segments can share access to a substation through adjacent electrified segments, not every deployed segment requires a direct grid connection. At $T\!=\!30$, 4 of the 12 selected DWC segments are supplied through adjacent links; by $T\!=\!32$, this increases to 18 of 30 segments (Fig.~\ref{fig:texas_transition}e). Additional DWC segments can therefore extend existing electrified corridors through shared grid access rather than requiring an independent distribution connection for each new segment. The rapid transition toward DWC thus reflects  these two reinforcing mechanisms, wherein higher charging throughput increases lifetime value as demand grows, while contiguous corridor connectivity reduces the additional grid infrastructure required as roadway electrification expands.

\subsubsection*{Traffic, construction cost, and grid access determine DWC and SCS deployment}
\label{sec:An interpretable screening tool}

\begin{figure}
    \centering
    \includegraphics[width=1\linewidth]{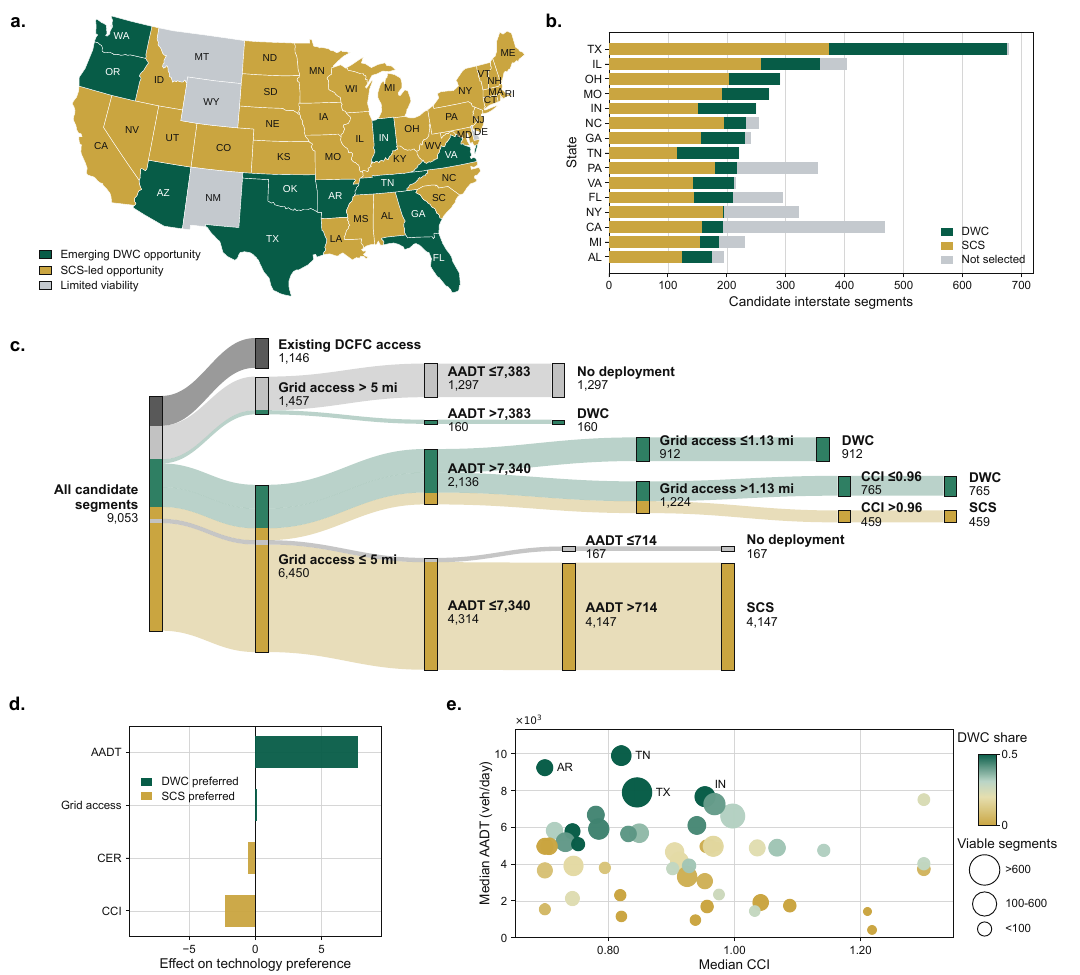}
    \caption{\textbf{Deployment outcomes reveal practical screening thresholds for DWC and SCS selection.}
    All panels are derived from the unconstrained national optimization with $T\!=\!32$ and baseline scenario parameters.
    \textbf{a.} State-level deployment archetypes. States are classified as having limited viability when fewer than 30\% of candidate segments receive either technology. Among the remaining states, emerging DWC opportunity is assigned when DWC accounts for at least 30\% of selected segments; all others are classified as SCS-led.
    \textbf{b.} DWC, SCS, and unselected candidate segments for the 15 states with the largest total deployment, ranked by total selections.
    \textbf{c.} Interpretable decision pathway summarizing optimized deployment outcomes. SCS candidates near existing DCFC stations are treated as already served and excluded from new SCS siting, while the corresponding roadway segments remain eligible for DWC. Direct grid connections are restricted to candidate substations within 5 miles (see Methods: \nameref{sec:GIS pipeline}). The resulting principal screening thresholds from the 32-year optimization are freight AADT of approximately 7,340-7,383 vehicles per day, substation distance of 1.13 miles, and a CCI of 0.96. 
    \textbf{d.} Standardized logistic-regression coefficients for technology preference among selected DWC and SCS segments. Positive coefficients favor DWC and negative coefficients favor SCS.
    \textbf{e.} State-level relationship between median AADT and median CCI across all candidate segments. Marker size indicates the number of viable segments and marker color indicates the DWC deployment share among selected DWC and SCS segments.}
    \label{fig:filteringtool}
\end{figure}

Having established when DWC becomes competitive and the mechanisms that drive its expansion, we next ask a broader planning question: \textit{what characteristics determine whether a roadway segment is better suited to DWC, SCS, or no new deployment?} We address this by evaluating the technology-selection optimization across the full candidate network and translating the resulting deployment patterns into quantitative screening rules. Although the resulting numerical thresholds are scenario-specific, the underlying relationships between traffic demand, construction cost, and grid access provide a generalizable basis for screening DWC and SCS opportunities in other roadway networks.

We find substantial geographic variation in the resulting deployment patterns (Fig.~\ref{fig:filteringtool}a-b). Within the US Interstate freight network, Texas, Washington, and Florida exhibit greater DWC opportunity, whereas several high deployment states, including California and Illinois, remain more SCS-oriented. These regional differences reflect local demand and construction conditions, with higher freight traffic and more favorable construction costs typically favoring DWC, while lower-demand regions more often favoring SCS or no new deployment.

These deployment patterns can be distilled into interpretable thresholds for DWC and SCS selection (Fig.~\ref{fig:filteringtool}c). Across the unconstrained 32-year scenario, the model selects 1,837 roadway segments for DWC and 4,606 locations for SCS. The resulting decision pathway identifies freight AADT of approximately 7,340-7,383 vehicles per day as a principal transition between SCS- and DWC-favored locations, together with a substation distance threshold of approximately 1.13 miles and a CCI threshold of 0.96. These thresholds quantify the local conditions under which the optimized technology choice shifts among DWC, SCS, and no new deployment.

Grid connectivity can further extend DWC viability beyond locations with direct substation access. We restrict direct grid connections to candidate substations within 5 miles of a roadway segment (see Methods: \nameref{sec:GIS pipeline}). Of the segments beyond this direct-access range, 1,457 are otherwise excluded, yet 160 remain viable for DWC through adjacent electrified segments. Shared substation access can therefore extend DWC deployment beyond locations that satisfy the direct grid-proximity requirement.

The regression analysis in Fig.~\ref{fig:filteringtool}d-e reinforces this hierarchy of drivers. AADT is the strongest positive predictor of DWC selection, while higher CCI favors SCS; grid access and CER have smaller effects once deployment is feasible. Traffic availability, therefore, creates the primary opportunity for DWC, while construction conditions and grid access determine where that opportunity can be economically realized. Together, these screening rules provide actionable guidance for identifying where DWC or SCS is most likely to be advantageous, and a generalizable framework (see Methods: \nameref{sec:si-Siting MILP}) that can be used to study deployment opportunities across roadway networks beyond the US freight system considered here.

\subsubsection*{Mixed DWC-SCS portfolios maximize lifetime value}\label{sec:Charging-demand allocation shapes portfolio composition and value}

\begin{figure}
    \centering
    \includegraphics[width=1\linewidth]{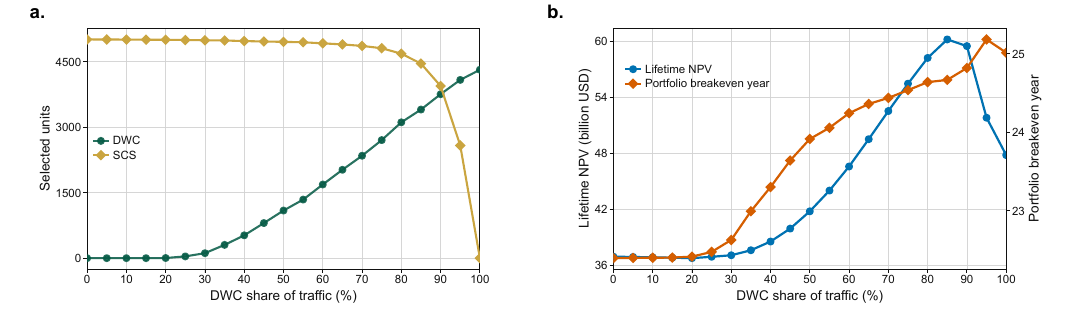}
    \caption{\textbf{EV traffic share reshapes portfolio composition and value.}
    Results are from the unconstrained national optimization with $T\!=\!32$ and all other parameters fixed at their baseline values. DWC traffic share is varied as a high-level proxy for DWC uptake and charging preference, with the remaining EV traffic assigned to SCS.
    \textbf{a.}  Numbers of selected DWC segments and SCS sites as the DWC traffic share increases from 0\% to 100\%. 
    \textbf{b.} Corresponding portfolio lifetime NPV and breakeven year. Portfolio value increases as the DWC traffic share rises, peaks at an intermediate share, and declines as the portfolio approaches the DWC-only endpoint, while breakeven is delayed as DWC participation increases.}
    \label{fig:traffic_split_variation}
\end{figure}

While demand growth and project lifetime can shift the portfolio toward DWC, the value of future infrastructure investments also depends on the extent to which EV charging demand is ultimately captured by DWC as the technology is adopted. To examine this effect, we parametrically vary the share of EV traffic allocated to DWC, using it as a high-level proxy for future DWC uptake and charging preference, with the remainder assigned to SCS. The portfolio response is strongly nonlinear (Fig.~\ref{fig:traffic_split_variation}a). DWC deployment begins to rise beyond roughly 25\% traffic share and expands steadily thereafter, while SCS deployment remains high until the DWC share approaches 80\%. Both technologies therefore coexist across a broad range of assumed DWC uptake, yielding mixed DWC-SCS portfolios over much of the adoption range. These mixed portfolios can produce greater long-term value than either technology alone (Fig.~\ref{fig:traffic_split_variation}b). Lifetime NPV peaks at approximately 85\% DWC traffic share, where about 3,400 DWC segments coexist with nearly 4,400 SCS sites. The resulting portfolio reaches USD~60.18 billion in lifetime value, 63\% above the SCS-only case and 26\% above the DWC-only case. Consistent with the spatial results above, DWC captures greater value in high-demand locations, while SCS remains valuable where lower utilization or local deployment conditions make roadway electrification less favorable. Increasing DWC participation raises portfolio value over most of the range but also delays breakeven; beyond the intermediate optimum, further displacement of SCS reduces lifetime value. These results show that the long-term value of roadway electrification goes beyond maximizing adoption of a single technology, highlighting the importance of how DWC and SCS are combined as charging demand is distributed between them.

\section*{Discussion}\label{sec:Discussion}

\noindent \textbf{Long-term planning and the  case for DWC.} The DWC-SCS trade-off illustrates a recurring infrastructure-investment problem in which established technologies can dominate near-term economics because of lower capital requirements, while capital-intensive alternatives become more valuable as utilization and network effects increase. The relevant question is therefore how the relative value of competing technologies evolves over the lifetime of the infrastructure, rather than which technology is cheaper today. For EV charging, our results show that portfolio composition depends on current technology costs, the growth of charging demand, and the duration over which infrastructure remains in service. This distinction is particularly important for charging investments designed to last for decades. Evaluating alternatives over short planning horizons or payback periods can favor SCS because its lower capital requirement enables earlier cost recovery, even when the same network would support substantially greater DWC deployment over a longer horizon as charging demand grows. DWC, by contrast, requires sufficient utilization and time for its higher charging throughput to offset its greater upfront cost. The planning horizon can thus materially alter which infrastructure is selected. These dynamics also create the potential for technological lock-in. Continued reliance on stationary charging can commit capital, charging locations, and grid connections to infrastructure designed around near-term demand, potentially reducing the value of later roadway electrification. Premature DWC deployment presents the opposite risk, stranding capital in corridors that do not achieve sufficient utilization. Long-term charging planning must therefore balance near-term capital recovery against the value of infrastructure under future demand rather than treating present-day technology economics as fixed over the asset lifetime.

\noindent \textbf{A generalized and transferable framework for charging infrastructure planning.} The optimization framework developed here (see Methods: \nameref{sec:si-Siting MILP}) provides a \textbf{\textit{full-fledged optimal siting and planning tool}} for jointly selecting DWC and SCS locations together with their supporting grid connections, taking into account spatially heterogeneous traffic demand, construction costs, electricity prices, existing charging infrastructure, and grid access. Rather than evaluating candidate technologies independently, the framework optimizes their combined deployment as a common investment portfolio while accounting for growing demand, competition for charging traffic, corridor connectivity, and lifetime economics. Beyond the optimized siting decisions themselves, the framework also produces interpretable thresholds that reveal the local conditions underlying technology selection. In the US Interstate freight network, traffic availability creates the primary opportunity for DWC, while construction conditions and grid access determine where that opportunity can be economically realized. Although the numerical thresholds obtained here are specific to the assumed network and scenarios, the planning framework is more general and can be adapted to other roadway networks and geographies by replacing the corresponding traffic, cost, grid, and existing-infrastructure inputs, enabling joint DWC-SCS siting and infrastructure planning under local conditions. The resulting thresholds provide an interpretable layer on top of the optimization, translating detailed deployment decisions into actionable insights for broader infrastructure planning.

\noindent \textbf{Differentiated electrification strategies and policy implications.}
Our results suggest that roadway electrification need not follow a uniform technology pathway. Mixed DWC-SCS portfolios remain valuable across a broad range of DWC uptake, while differences in traffic demand, construction conditions, and grid access create distinct regional opportunities for the two technologies. DWC also changes the structure of grid investment, shifting from many smaller, site-specific connections toward larger corridor-level connections that can be shared across contiguous electrified segments. Regional charging strategies should therefore account for the scale and location of charging demand together with the way alternative technology portfolios interact with local transportation and grid infrastructure. Our US results are also consistent with several priorities emerging in current planning efforts. Texas, which shows strong DWC opportunity in our analysis, is actively evaluating medium- and heavy-duty corridor charging needs and grid coordination in its freight-electrification assessment \cite{txdot2024rider48}. Tennessee similarly identifies high-freight corridors as important to statewide electrification planning \cite{tdec2026freight}. California, by contrast, remains more SCS-led in our results, consistent with Bay Area planning that anticipates 390-580 MW of stationary medium- and heavy-duty charging demand by 2035 \cite{rmi2026bayarea}. These examples show that the traffic and infrastructure conditions highlighted by our model are also central to current planning efforts and can produce substantially different regional infrastructure priorities. As national electrified-roadway initiatives and pilot programs (e.g., ASPIRE \cite{aspire2026center}) advance toward larger-scale deployment, our framework can help identify highway corridors that warrant detailed technical and financial evaluation. Our results further suggest that corridor-level interconnection planning may be especially important for DWC because shared grid connections contribute to its economic advantage. More broadly, spatially differentiated planning can help identify where DWC offers the strongest opportunity, where stationary charging remains more appropriate, and where coordinated transportation and grid investment is most consequential.

In summary, these results provide a basis for rethinking roadway electrification as a coordinated, long-term infrastructure transition shaped by evolving demand, technology choice, and grid conditions.

\section*{Methods}\label{sec:methods}

\subsection*{Overview}\label{sec:Overview}
Our methodology integrates a geospatial data-processing workflow with a mixed-integer linear programming (MILP)-based siting optimization framework for the joint placement of DWC and SCS infrastructure. We process GIS data describing the US Interstate freight network and associated traffic, charging infrastructure, cost, and grid attributes in ArcGIS Pro 3.6.0 to construct the MILP inputs. These inputs are then supplied to the siting framework, implemented in Python and solved using Gurobi Optimizer 13.0.1. The resulting optimal deployment decisions are subsequently mapped back onto the network for visualization and spatial analysis.

\subsection*{GIS workflow}\label{sec:GIS pipeline}
The GIS workflow converts the US Interstate freight network into a spatially resolved set of roadway candidates for siting optimization, with all processing performed in the USA Contiguous Albers Equal Area Conic projected coordinate system. We first assign CER and CCI attributes to the original Interstate features using spatial joins. The roadway features are then dissolved by route and divided into approximately 5 mile segments, with residual segments removed. We transfer the source attributes to the segmented network using spatial joins, retaining maximum intersecting values and replacing missing traffic values with the corresponding dataset medians and missing CER and CCI values with the corresponding dataset averages. Candidate SCS sites are represented using Interstate ramps, excluding those already served by existing DCFC infrastructure and their associated charging demand. We represent grid accessibility by identifying substations within a 5 mile radius of roadway and SCS candidates. The resulting segment attributes, candidate charging sites, substation connections, and roadway-adjacency relationships form the geospatial inputs to the MILP. Exact GIS operations and input formatting are provided in Supplementary Note~\ref{sec:si-GIS Workflow}.

\subsection*{Joint DWC and SCS investment planning and siting framework}\label{sec:si-Siting MILP}

We jointly select DWC segments, SCS sites, and supporting grid infrastructure to maximize the lifetime NPV of the overall charging portfolio, which we refer to as its lifetime value. Charging revenue is generated from the energy delivered to EVs, and portfolio NPV is defined as the discounted value of annual charging revenues after accounting for electricity and fixed operation and maintenance (O\&M) costs, less upfront capital expenditures.

\subsubsection*{a. Lifetime growth and present-value parameters}\label{subsec:si-Lifetime growth and present-value parameters}

We evaluate the charging portfolio over a lifetime of $T$ years, with $\mathcal{Y}=\{1,\ldots,T\}$ and $y=1$ denoting the base year. We discount future annual charging revenues and operating costs using the annual discount rate $\rho$. We assume annual revenues and operating costs to occur at the end of each project year. Traffic, EV adoption, charging prices, commercial electricity rates, and fixed O\&M costs are allowed to grow over the project lifetime.

Let $g^{\mathrm{AADT}}$ denote the annual traffic growth rate and $g^{\mathrm{EV}}$ the annual EV-adoption growth rate. Their compounded effect gives the annual charging-demand growth rate:
\begin{equation}
g^{\mathrm{d}}=(1+g^{\mathrm{AADT}})(1+g^{\mathrm{EV}})-1.
\label{eq:si-demand_growth}
\end{equation}

The corresponding charging-demand multiplier in project year $y$ is:
\begin{equation}
\gamma^{\mathrm{d}}_y=(1+g^{\mathrm{d}})^{y-1},
\qquad y\in\mathcal{Y},
\label{eq:si-demand_multiplier}
\end{equation}
with $\gamma^{\mathrm{d}}_1=1$ in the base year.

Charging revenues and electricity costs grow through the combined effects of charging demand growth and the corresponding rate escalation over time. Let $g^{\mathrm{p,DWC}}$ and $g^{\mathrm{p,SCS}}$ denote the annual growth rates of DWC and SCS retail charging prices, and let $g^{\mathrm{c,CER}}$ denote the annual CER growth rate. The resulting effective growth rates are:
\begin{equation}
\begin{aligned}
g^{\mathrm{rev,DWC}}
&=(1+g^{\mathrm{d}})(1+g^{\mathrm{p,DWC}})-1,\\
g^{\mathrm{rev,SCS}}
&=(1+g^{\mathrm{d}})(1+g^{\mathrm{p,SCS}})-1,\\
g^{\mathrm{c,elec}}
&=(1+g^{\mathrm{d}})(1+g^{\mathrm{c,CER}})-1.
\end{aligned}
\label{eq:si-effective_growth_rates}
\end{equation}

Fixed O\&M costs are independent of charging demand and grow annually at the rate $g^{\mathrm{c,om}}$.

For any annual charging revenue or operating cost with base year value $x_1$, and annual growth rate $g$, its discounted lifetime value is represented by $x_1\Gamma(g)$ \cite{PValue}, where:
\begin{equation}
\Gamma(g)=
\sum_{y=1}^{T}
\frac{(1+g)^{y-1}}{(1+\rho)^y}.
\label{eq:si-pv_multiplier}
\end{equation}

Accordingly, we convert the base year annual DWC revenue to its discounted lifetime value using $\Gamma(g^{\mathrm{rev,DWC}})$, SCS revenue using $\Gamma(g^{\mathrm{rev,SCS}})$, electricity costs using $\Gamma(g^{\mathrm{c,elec}})$, and fixed O\&M costs using $\Gamma(g^{\mathrm{c,om}})$. Upfront capital costs are incurred only once at project initiation and are therefore included directly without this lifetime-value conversion.

\subsubsection*{b. Charging demand}\label{subsec:si-Charging demand}

We construct charging demand for the base year at the roadway-segment and candidate-SCS level, and subsequently scale it  over the project lifetime using the demand-growth multiplier $\gamma^{\mathrm{d}}_y$ defined in equation \eqref{eq:si-demand_multiplier}. Let $\mathcal{I}$ denote the set of $n$ roadway segments and $\mathcal{R}$ the set of $m$ candidate SCS sites retained for optimization. Throughout, \(\mathbf{1}_q\) and \(\mathbf{0}_q\) denote \(q\)-dimensional vectors of ones and zeros, respectively, \((\cdot)^\top\) denotes transpose, \(\operatorname{diag}(\cdot)\) forms a diagonal matrix from a vector, and \(\odot\) and \(\oslash\) denote elementwise multiplication and division, respectively.

For the roadway network, let $\mathbf{a}\in\mathbb{R}^{n}_{+}$ denote the vector of base year AADT values, where $a_i$ is the AADT of segment $i$. We distribute daily traffic over a normalized 24-hour profile $\boldsymbol{\omega}\in\mathbb{R}^{24}_{+}$, where $\omega_h$ is the fraction of daily traffic occurring during hour $h$. Let $\rho^{\mathrm{EV}}_1$ denote the base year fraction of traffic that is electric, while $\eta^{\mathrm{DWC}}$ and $\eta^{\mathrm{SCS}}$ denote the exogenous shares of EV traffic assigned to DWC and SCS charging, respectively.

For DWC, let $\mathbf{l}\in\mathbb{R}^{n}_{+}$ and $\mathbf{v}\in\mathbb{R}^{n}_{+}$ denote the roadway-segment length and travel-speed vectors, and let $p^{\mathrm{DWC}}$ denote the rated EV-side DWC receiver power. For segment $i$, the travel time is $l_i/v_i$, such that $p^{\mathrm{DWC}}l_i/v_i$ gives the energy received by an EV assigned to DWC charging while traversing segment $i$ if it were electrified. We use the EV-side receiver rating rather than propulsion power, representing charging as constrained by the EV charging hardware. Because $\mathbf{1}_{24}^{\top}\boldsymbol{\omega}=1$ and the average segment speed is assumed constant across hours, the hourly terms collapse in the annual demand calculation:
\begin{equation}
\mathbf{e}^{\mathrm{traff,DWC}}_{1}
=
365
\rho^{\mathrm{EV}}_1
\eta^{\mathrm{DWC}}
p^{\mathrm{DWC}}
\operatorname{diag}\!\left(\mathbf{l}\oslash\mathbf{v}\right)
\mathbf{a},
\label{eq:si-dwc_demand_vector}
\end{equation}
where $\mathbf{e}^{\mathrm{traff,DWC}}_{1}\in\mathbb{R}^{n}_{+}$ contains the annual traffic-derived DWC charging demand for all roadway segments.

For SCS demand, let $\mathcal{H}_r\subseteq\mathcal{I}$ denote the set of roadway segments whose traffic is associated with candidate site $r$. We define the host-assignment matrix $\mathbf{G}\in\{0,1\}^{m\times n}$ such that $G_{r,i}=1$ when $i\in\mathcal{H}_r$ and $G_{r,i}=0$ otherwise. Hence, $\mathbf{G}\mathbf{a}$ gives the aggregate AADT associated with each candidate SCS site. Let $e^{\mathrm{req}}$ denote the energy delivered during one SCS charging event. The base year traffic-derived SCS charging demand is then:
\begin{equation}
\mathbf{e}^{\mathrm{traff,SCS}}_{1}
=
365
\rho^{\mathrm{EV}}_1
\eta^{\mathrm{SCS}}
e^{\mathrm{req}}
\mathbf{G}\mathbf{a},
\label{eq:si-scs_demand_vector}
\end{equation}
where $\mathbf{e}^{\mathrm{traff,SCS}}_{1}\in\mathbb{R}^{m}_{+}$ contains the annual traffic-derived SCS charging demand across candidate sites. Thus, DWC demand is evaluated at the roadway-segment level, whereas SCS demand aggregates traffic from all roadway segments associated with each candidate site. Traffic-derived charging demand in project year $y\in\mathcal{Y}$ is then obtained as:
\begin{equation}
\begin{aligned}
\mathbf{e}^{\mathrm{traff,DWC}}_{y}
&=
\gamma^{\mathrm{d}}_y
\mathbf{e}^{\mathrm{traff,DWC}}_{1},\\
\mathbf{e}^{\mathrm{traff,SCS}}_{y}
&=
\gamma^{\mathrm{d}}_y
\mathbf{e}^{\mathrm{traff,SCS}}_{1},
\end{aligned}
\qquad y\in\mathcal{Y}.
\label{eq:si-traffic_demand_year}
\end{equation}

\subsubsection*{c. System revenues and costs}\label{subsec:si-System revenues and costs}

We evaluate the economic contribution of each deployment using base year charging revenue, electricity costs, fixed O\&M costs, and upfront capital costs. The recurring operating costs comprise electricity costs and fixed O\&M costs, whose lifetime present values are calculated using the growth and discounting factors defined above.\\

\noindent \textbf{Base year charging revenue:}
Let $\mathbf{e}^{\mathrm{del,DWC}}_{1}\in\mathbb{R}^{n}_{+}$ and $\mathbf{e}^{\mathrm{del,SCS}}_{1}\in\mathbb{R}^{m}_{+}$ denote the base year delivered-energy decision vectors for DWC and SCS, respectively. Let $c^{\mathrm{rev,DWC}}$ and $c^{\mathrm{rev,SCS}}$ denote the corresponding base year retail charging prices. The total base year revenues are:
\begin{equation}
R^{\mathrm{DWC}}_{1}
=
c^{\mathrm{rev,DWC}}
\mathbf{1}_{n}^{\top}
\mathbf{e}^{\mathrm{del,DWC}}_{1},
\label{eq:si-revenue_dwc}
\end{equation}
\begin{equation}
R^{\mathrm{SCS}}_{1}
=
c^{\mathrm{rev,SCS}}
\mathbf{1}_{m}^{\top}
\mathbf{e}^{\mathrm{del,SCS}}_{1}.
\label{eq:si-revenue_scs}
\end{equation}

The DWC and SCS revenue subsequently grow at
$g^{\mathrm{rev,DWC}}$ and $g^{\mathrm{rev,SCS}}$, respectively, as defined in equation \eqref{eq:si-effective_growth_rates}.\\

\noindent \textbf{Regional cost scaling:}
Let $\mathbf{c}^{\mathrm{CER,DWC}}\in\mathbb{R}^{n}_{+}$ and
$\mathbf{c}^{\mathrm{CCI,DWC}}\in\mathbb{R}^{n}_{+}$ denote the segment-level
CER and CCI vectors, respectively.
Candidate SCS sites inherit AADT-weighted CER and CCI values from their associated
roadway segments. Using the host-assignment matrix $\mathbf{G}$ defined previously,
for candidate sites with positive aggregate AADT:
\begin{equation}
\begin{aligned}
\mathbf{c}^{\mathrm{CER,SCS}}
&=
\operatorname{diag}\!\left(\mathbf{G}\mathbf{a}\right)^{-1}
\mathbf{G}\operatorname{diag}(\mathbf{a})
\mathbf{c}^{\mathrm{CER,DWC}},\\
\mathbf{c}^{\mathrm{CCI,SCS}}
&=
\operatorname{diag}\!\left(\mathbf{G}\mathbf{a}\right)^{-1}
\mathbf{G}\operatorname{diag}(\mathbf{a})
\mathbf{c}^{\mathrm{CCI,DWC}}.
\end{aligned}
\label{eq:si-scs_weighted_attributes}
\end{equation}

Here, $\mathbf{c}^{\mathrm{CER,SCS}}\in\mathbb{R}^{m}_{+}$ and
$\mathbf{c}^{\mathrm{CCI,SCS}}\in\mathbb{R}^{m}_{+}$ contain the AADT-weighted
CER and CCI values associated with the candidate SCS sites. For a candidate
whose associated segments have zero aggregate AADT, we assign the CER and CCI
of its representative host segment, consistent with the implementation.\\

\noindent \textbf{Electricity costs:}
Using the CER vectors defined above, the base year electricity costs are:
\begin{equation}
c^{\mathrm{elec,DWC}}_{1}
=
\left(\mathbf{c}^{\mathrm{CER,DWC}}\right)^{\top}
\mathbf{e}^{\mathrm{del,DWC}}_{1},
\label{eq:si-electricity_cost_dwc}
\end{equation}
\begin{equation}
c^{\mathrm{elec,SCS}}_{1}
=
\left(\mathbf{c}^{\mathrm{CER,SCS}}\right)^{\top}
\mathbf{e}^{\mathrm{del,SCS}}_{1}.
\label{eq:si-electricity_cost_scs}
\end{equation}

CER scales only the energy-procurement component and is not multiplied by the regional CCI. Annual electricity costs for both technologies grow over the project lifetime at the effective rate $g^{\mathrm{c,elec}}$.\\

\noindent \textbf{Fixed O\&M costs:}
Let $\boldsymbol{\delta}^{\mathrm{DWC}}\in\{0,1\}^{n}$ and $\boldsymbol{\delta}^{\mathrm{SCS}}\in\{0,1\}^{m}$ denote the binary DWC and SCS deployment vectors, respectively. Let $c^{\mathrm{om,DWC}}$ and $c^{\mathrm{om,SCS}}$ denote the corresponding base annual fixed O\&M costs per deployed DWC segment and SCS station. For both technologies, we assume the base annual fixed O\&M cost to equal 1.5\% of the corresponding installed infrastructure and grid-connection cost \cite{Haddad2022FinancialViability}. Regional CCI values scale these costs to account for spatial differences in infrastructure cost. The base year fixed O\&M costs are therefore:
\begin{equation}
c^{\mathrm{fix,DWC}}_{1}
=
c^{\mathrm{om,DWC}}
\left(\mathbf{c}^{\mathrm{CCI,DWC}}\right)^{\top}
\boldsymbol{\delta}^{\mathrm{DWC}},
\label{eq:si-fixed_cost_dwc}
\end{equation}
\begin{equation}
c^{\mathrm{fix,SCS}}_{1}
=
c^{\mathrm{om,SCS}}
\left(\mathbf{c}^{\mathrm{CCI,SCS}}\right)^{\top}
\boldsymbol{\delta}^{\mathrm{SCS}}.
\label{eq:si-fixed_cost_scs}
\end{equation}

The fixed O\&M costs depend only on deployed infrastructure and not on delivered charging energy. They grow over the project lifetime at the annual rate $g^{\mathrm{c,om}}$.\\

\noindent\textbf{Upfront capital costs:}
We account for the cost of deploying DWC segments and SCS stations, together with the infrastructure needed to supply them with power. This supporting infrastructure includes adjacent links between contiguous DWC segments and distribution lines connecting DWC segments or SCS stations directly to substations. Each direct substation connection incurs a distribution-line construction cost based on its length and a grid-connection cost based on the power capacity required to supply the charging infrastructure. All capital costs are incurred once at project initiation and scaled by the corresponding regional CCI.

Each adjacent link connects two neighboring DWC segments and can
supply power in either direction. In the optimization, we represent these two possible flow directions separately: $(i,j)$ denotes flow from segment $i$ to segment $j$, and $(j,i)$ denotes the reverse direction. These directional representations, termed directed arcs, describe the same physical adjacent link. Let $\mathcal{A}$ denote the set of these directed arcs, with
$|\mathcal{A}|=n^{\mathrm{adj}}$. The binary variable
$\delta^{\mathrm{adj}}_k$ indicates whether the adjacent link is
constructed with the flow direction represented by arc $k$.
The optimization permits at most one of the two directions to be
selected for each physical adjacent link.

We scale the construction cost of an adjacent link by the mean
CCI of the two segments it
connects. To calculate the mean CCI for each candidate
link, we define the endpoint-incidence matrix
$\mathbf{L}^{\mathrm{adj}}\!\in\!\{0,1\}^{n^{\mathrm{adj}}\times n}$.
For an arc $k$ connecting segments $i$ and $j$, row $k$ contains
ones in columns $i$ and $j$ and zeros elsewhere. The corresponding
entry of
$\frac{1}{2}\mathbf{L}^{\mathrm{adj}}\mathbf{c}^{\mathrm{CCI,DWC}}$
is therefore
$\frac{1}{2}(c_i^{\mathrm{CCI,DWC}}+c_j^{\mathrm{CCI,DWC}})$.
The rows for $(i,j)$ and $(j,i)$ are identical because both
directions represent the same physical link and use the same
regional CCI scaling.

Let $\mathcal{K}^{\mathrm{DWC}}$ and $\mathcal{K}^{\mathrm{SCS}}$
denote the sets of candidate substation connections for DWC segments
and SCS sites, with $k^{\mathrm{DWC}}=|\mathcal{K}^{\mathrm{DWC}}|$
and $k^{\mathrm{SCS}}=|\mathcal{K}^{\mathrm{SCS}}|$ denoting the corresponding numbers of candidate connections.
The matrices
$\mathbf{T}^{\mathrm{DWC}}\in\{0,1\}^{n\times k^{\mathrm{DWC}}}$
and
$\mathbf{T}^{\mathrm{SCS}}\in\{0,1\}^{m\times k^{\mathrm{SCS}}}$
identify the DWC segment or SCS site supplied by each connection. Each column represents one connection and contains a
single one in the row of the segment or site it supplies, with zeros
elsewhere. Thus,
$(\mathbf{T}^{\mathrm{DWC}})^{\top}\mathbf{c}^{\mathrm{CCI,DWC}}$
and
$(\mathbf{T}^{\mathrm{SCS}})^{\top}\mathbf{c}^{\mathrm{CCI,SCS}}$
give the CCI used to scale each connection's construction and
capacity costs.

Let $\mathbf{d}^{\mathrm{DWC}}\in\mathbb{R}_{+}^{k^{\mathrm{DWC}}}$
and $\mathbf{d}^{\mathrm{SCS}}\in\mathbb{R}_{+}^{k^{\mathrm{SCS}}}$
contain the distribution-line lengths for these connections.
The binary vectors
$\boldsymbol{\delta}^{\mathrm{sub,DWC}}\in\{0,1\}^{k^{\mathrm{DWC}}}$
and
$\boldsymbol{\delta}^{\mathrm{sub,SCS}}\in\{0,1\}^{k^{\mathrm{SCS}}}$
indicate which connections are constructed, while
$\mathbf{p}^{\mathrm{tap,DWC}}\in\mathbb{R}_{+}^{k^{\mathrm{DWC}}}$
and
$\mathbf{p}^{\mathrm{tap,SCS}}\in\mathbb{R}_{+}^{k^{\mathrm{SCS}}}$
give the power supplied through them (physical power injections).

The base costs are $c^{\mathrm{DWC}}$ for infrastructure per DWC
segment, $c^{\mathrm{int}}$ for internal electrical lines per DWC
segment, $c^{\mathrm{adj}}$ per adjacent link,
$c^{\mathrm{ext}}$ per unit distribution-line length, and
$c^{\mathrm{conn}}$ per unit connected power capacity.
Let \(n^{\mathrm{SCS}}\) denote the number of charging ports per SCS station and \(c^{\mathrm{port}}\) the infrastructure cost per port. The infrastructure cost per SCS station is then \(c^{\mathrm{SCS}}=n^{\mathrm{SCS}}c^{\mathrm{port}}\).
The total upfront capital cost is:
\begin{equation}
\begin{aligned}
c^{\mathrm{cap}}
&={}
\left(c^{\mathrm{DWC}}+c^{\mathrm{int}}\right)
\left(\mathbf{c}^{\mathrm{CCI,DWC}}\right)^{\top}
\boldsymbol{\delta}^{\mathrm{DWC}}
+c^{\mathrm{adj}}
\left[
\frac{1}{2}
\mathbf{L}^{\mathrm{adj}}
\mathbf{c}^{\mathrm{CCI,DWC}}
\right]^{\top}
\boldsymbol{\delta}^{\mathrm{adj}}
\\[2mm]
&+
c^{\mathrm{ext}}
\left[
\left(
(\mathbf{T}^{\mathrm{DWC}})^{\top}
\mathbf{c}^{\mathrm{CCI,DWC}}
\right)
\odot
\mathbf{d}^{\mathrm{DWC}}
\right]^{\top}\!
\boldsymbol{\delta}^{\mathrm{sub,DWC}}
+
c^{\mathrm{conn}}
\left[
(\mathbf{T}^{\mathrm{DWC}})^{\top}
\mathbf{c}^{\mathrm{CCI,DWC}}
\right]^{\top}\!
\mathbf{p}^{\mathrm{tap,DWC}}
\\[2mm]
&+
c^{\mathrm{SCS}}
\left(\mathbf{c}^{\mathrm{CCI,SCS}}\right)^{\top}
\boldsymbol{\delta}^{\mathrm{SCS}}
+
c^{\mathrm{ext}}
\left[
\left(
(\mathbf{T}^{\mathrm{SCS}})^{\top}
\mathbf{c}^{\mathrm{CCI,SCS}}
\right)
\odot
\mathbf{d}^{\mathrm{SCS}}
\right]^{\top}\!
\boldsymbol{\delta}^{\mathrm{sub,SCS}}
+
c^{\mathrm{conn}}\!
\left[
(\mathbf{T}^{\mathrm{SCS}})^{\top}
\mathbf{c}^{\mathrm{CCI,SCS}}
\right]^{\top}\!
\mathbf{p}^{\mathrm{tap,SCS}} .
\end{aligned}
\label{eq:si-total_capex}
\end{equation}

The terms in equation \eqref{eq:si-total_capex} respectively represent DWC infrastructure and internal wiring, DWC adjacent links, DWC distribution-line construction, DWC capacity-dependent grid-connection costs, SCS infrastructure, SCS distribution-line construction, and SCS capacity-dependent grid-connection costs. Because $c^{\mathrm{cap}}$ is incurred at project initiation, we do not multiply it by a lifetime growth or present-value factor.

\subsubsection*{d. Joint DWC and SCS siting optimization problem}\label{subsec:si-Joint DWC--SCS deployment optimization}

With the demand, revenue, and cost components defined above, we formulate the joint DWC--SCS deployment problem as an MILP. The model selects DWC segments and SCS sites together with the supporting electrical infrastructure required to supply the deployed charging assets. The resulting portfolio maximizes lifetime NPV subject to an optional deployment constraint, network-connectivity, power-flow, and charging-capacity constraints.

We begin by defining the network matrix used to enforce flow conservation. While $\mathbf{L}^{\mathrm{adj}}$ identifies the two roadway segments connected by each directed arc, the signed incidence matrix $\mathbf{B}\in\{-1,0,1\}^{n\times n^{\mathrm{adj}}}$ additionally encodes arc direction. For an arc $k=(j,i)$ directed from segment $j$ to segment $i$, $B_{j,k}=-1$, $B_{i,k}=1$, and all other entries in column $k$ are zero. Let $\mathbf{u}^{\mathrm{adj}}\in\mathbb{R}^{n^{\mathrm{adj}}}_{+}$ denote the virtual flow along the directed adjacency arcs and $\mathbf{f}\in\mathbb{R}^{n^{\mathrm{adj}}}_{+}$ the corresponding physical power flow. Multiplication by $\mathbf{B}$ therefore gives the net incoming virtual or physical flow at each roadway segment.

Let $n^{\mathrm{sub}}$ denote the number of substations represented in the DWC connectivity network. The matrix $\mathbf{Q}\in\{0,1\}^{n^{\mathrm{sub}}\times k^{\mathrm{DWC}}}$ maps each candidate DWC substation connection to its source substation. Let $\mathbf{u}^{\mathrm{sub}}\in\mathbb{R}^{k^{\mathrm{DWC}}}_{+}$ denote the virtual flow through the candidate DWC substation connections, and let $\mathbf{w}\in\mathbb{R}^{n^{\mathrm{sub}}}_{+}$ denote the virtual supply originating from the substations. Let $n^{\mathrm{link}}$ denote the number of neighboring roadway-segment pairs without regard to direction. The matrix $\mathbf{U}\in\{0,1\}^{n^{\mathrm{link}}\times n^{\mathrm{adj}}}$ groups the two opposite directed arcs representing each physical adjacent link.

Let $p^{\mathrm{coil}}$ denote the installed DWC power density per unit roadway length and $p^{\mathrm{SCS}}$ the rated power per SCS port. The rated charging capacities of the installed DWC segments and SCS stations are:
\begin{equation}
\bar{\mathbf{p}}^{\mathrm{DWC}}
=
p^{\mathrm{coil}}\mathbf{l},
\qquad
p^{\mathrm{SCS,stn}}
=
n^{\mathrm{SCS}}p^{\mathrm{SCS}}.
\label{eq:si-installed_power_capacity}
\end{equation}

These infrastructure ratings specify the total power available at a DWC segment or SCS station; the EV-side receiver power $p^{\mathrm{DWC}}$ specifies the power received by an individual EV. The corresponding annual charging-energy capacities are:
\begin{equation}
\mathbf{e}^{\mathrm{cap,DWC}}
=
1000H^{\mathrm{op}}\bar{\mathbf{p}}^{\mathrm{DWC}},
\qquad
e^{\mathrm{cap,SCS}}
=
1000\mu^{\mathrm{SCS}}H^{\mathrm{op}}p^{\mathrm{SCS,stn}},
\label{eq:si-physical_energy_capacity}
\end{equation}
where $H^{\mathrm{op}}$ is the annual operating duration and $\mu^{\mathrm{SCS}}$ is the SCS utilization factor. The factor $1000$ converts MW to kW when annual energy is expressed in kWh. These annual capacities bound the energy that can be delivered by the installed infrastructure; actual delivered energy is a decision variable constrained by both traffic-derived charging demand and these capacities.

Let $\Pi$ denote the maximum combined number of newly deployed DWC segments and SCS stations. The implementation uses two Big-$M$ parameters. The virtual-flow bound is based on this deployment limit, while the physical-flow bound is the aggregate installed DWC power requirement:

\begin{equation}
M^{\mathrm{virt}}=\Pi+1,
\qquad
M^{\mathrm{pow}}
=
\mathbf{1}_{n}^{\top}
\bar{\mathbf{p}}^{\mathrm{DWC}}.
\label{eq:si-big_m_values}
\end{equation}

For scenarios without an explicit deployment-count limit, \(M^{\mathrm{virt}}\) is instead assigned a sufficiently large bound based on the maximum feasible DWC deployments. Collecting the decision variables used throughout the formulation, let:
\begin{equation}
\mathcal{V}
=
\left\{
\boldsymbol{\delta}^{\mathrm{DWC}},
\boldsymbol{\delta}^{\mathrm{SCS}},
\boldsymbol{\delta}^{\mathrm{adj}},
\boldsymbol{\delta}^{\mathrm{sub,DWC}},
\boldsymbol{\delta}^{\mathrm{sub,SCS}},
\mathbf{u}^{\mathrm{adj}},
\mathbf{u}^{\mathrm{sub}},
\mathbf{w},
\mathbf{f},
\mathbf{p}^{\mathrm{tap,DWC}},
\mathbf{p}^{\mathrm{tap,SCS}},
\mathbf{e}^{\mathrm{del,DWC}}_{1},
\mathbf{e}^{\mathrm{del,SCS}}_{1}
\right\}.
\label{eq:si-decision_variable_set}
\end{equation}

Let $Z$ denote the lifetime portfolio NPV. Using these definitions, we define the joint deployment problem as:
\begin{subequations}
\label{eq:si-joint_optimization}

\begin{equation}
\begin{aligned}
\underset{\mathcal{V}}{\max}\quad Z
={}&
\Gamma\!\left(g^{\mathrm{rev,DWC}}\right)R^{\mathrm{DWC}}_{1}
+
\Gamma\!\left(g^{\mathrm{rev,SCS}}\right)R^{\mathrm{SCS}}_{1}
-
\Gamma\!\left(g^{\mathrm{c,elec}}\right)
\left(
c^{\mathrm{elec,DWC}}_{1}
+
c^{\mathrm{elec,SCS}}_{1}
\right)\\
&-
\Gamma\!\left(g^{\mathrm{c,om}}\right)
\left(
c^{\mathrm{fix,DWC}}_{1}
+
c^{\mathrm{fix,SCS}}_{1}
\right)
-
c^{\mathrm{cap}},
\end{aligned}
\label{eq:si-joint_objective}
\end{equation}

\vspace{-0.3em}
\noindent\hspace{0.145\textwidth}\textbf{subject to:}
\vspace{0.45em}

\noindent\hspace{0.165\textwidth}\textit{Deployment constraint:}
\vspace{-0.5em}
\begin{align}
\mathbf{1}_{n}^{\top}\boldsymbol{\delta}^{\mathrm{DWC}}
+
\mathbf{1}_{m}^{\top}\boldsymbol{\delta}^{\mathrm{SCS}}
&\leq
\Pi,
\label{eq:si-deployment_limit}
\end{align}

\vspace{-0.6em}
\noindent\hspace{0.165\textwidth}\textit{Virtual connectivity constraints:}
\vspace{-0.5em}
\begin{align}
\mathbf{B}\mathbf{u}^{\mathrm{adj}}
+
\mathbf{T}^{\mathrm{DWC}}\mathbf{u}^{\mathrm{sub}}
&=
\boldsymbol{\delta}^{\mathrm{DWC}},
\label{eq:si-virtual_segment_balance}\\
\mathbf{w}
-
\mathbf{Q}\mathbf{u}^{\mathrm{sub}}
&=
\mathbf{0}_{n^{\mathrm{sub}}},
\label{eq:si-virtual_substation_balance}\\
\mathbf{1}_{n^{\mathrm{sub}}}^{\top}\mathbf{w}
&=
\mathbf{1}_{n}^{\top}\boldsymbol{\delta}^{\mathrm{DWC}},
\label{eq:si-virtual_total_supply}
\end{align}

\vspace{-0.6em}
\noindent\hspace{0.165\textwidth}\textit{DWC physical power-flow constraint:}
\vspace{-0.5em}
\begin{align}
\mathbf{B}\mathbf{f}
+
\mathbf{T}^{\mathrm{DWC}}\mathbf{p}^{\mathrm{tap,DWC}}
&=
\operatorname{diag}\!\left(
\bar{\mathbf{p}}^{\mathrm{DWC}}
\right)
\boldsymbol{\delta}^{\mathrm{DWC}},
\label{eq:si-dwc_power_balance}
\end{align}

\vspace{-0.6em}
\noindent\hspace{0.165\textwidth}\textit{DWC substation-connection constraints:}
\vspace{-0.5em}
\begin{alignat}{2}
\mathbf{u}^{\mathrm{sub}}
&\leq
M^{\mathrm{virt}}\boldsymbol{\delta}^{\mathrm{sub,DWC}},
\qquad&
\mathbf{p}^{\mathrm{tap,DWC}}
&\leq
M^{\mathrm{pow}}\boldsymbol{\delta}^{\mathrm{sub,DWC}},
\label{eq:si-dwc_substation_coupling}\\
\boldsymbol{\delta}^{\mathrm{sub,DWC}}
&\leq
\left(\mathbf{T}^{\mathrm{DWC}}\right)^{\top}
\boldsymbol{\delta}^{\mathrm{DWC}},
\label{eq:si-dwc_substation_deployment_coupling}
\end{alignat}

\vspace{-0.6em}
\noindent\hspace{0.165\textwidth}\textit{DWC adjacent-link constraints:}
\vspace{-0.5em}
\begin{alignat}{2}
\mathbf{u}^{\mathrm{adj}}
&\leq
M^{\mathrm{virt}}\boldsymbol{\delta}^{\mathrm{adj}},
\qquad&
\mathbf{f}
&\leq
M^{\mathrm{pow}}\boldsymbol{\delta}^{\mathrm{adj}},
\label{eq:si-dwc_adjacency_flow_coupling}\\
2\boldsymbol{\delta}^{\mathrm{adj}}
&\leq
\mathbf{L}^{\mathrm{adj}}
\boldsymbol{\delta}^{\mathrm{DWC}},
\qquad&
\mathbf{U}\boldsymbol{\delta}^{\mathrm{adj}}
&\leq
\mathbf{1}_{n^{\mathrm{link}}},
\label{eq:si-dwc_adjacency_logic}
\end{alignat}

\vspace{-0.6em}
\noindent\hspace{0.165\textwidth}\textit{SCS connection constraints:}
\vspace{-0.5em}
\begin{alignat}{2}
\mathbf{p}^{\mathrm{tap,SCS}}
&\leq
M^{\mathrm{pow}}\boldsymbol{\delta}^{\mathrm{sub,SCS}},
\qquad&
\boldsymbol{\delta}^{\mathrm{sub,SCS}}
&\leq
\left(\mathbf{T}^{\mathrm{SCS}}\right)^{\top}
\boldsymbol{\delta}^{\mathrm{SCS}},
\label{eq:si-scs_connection_coupling}\\
\mathbf{T}^{\mathrm{SCS}}\mathbf{p}^{\mathrm{tap,SCS}}
&=
p^{\mathrm{SCS,stn}}\boldsymbol{\delta}^{\mathrm{SCS}},
\qquad&
\mathbf{T}^{\mathrm{SCS}}\boldsymbol{\delta}^{\mathrm{sub,SCS}}
&\leq
\mathbf{1}_{m},
\label{eq:si-scs_power_connection}
\end{alignat}

\vspace{-0.6em}
\noindent\hspace{0.165\textwidth}\textit{Charging-energy constraints:}
\vspace{-0.5em}
\begin{alignat}{2}
\mathbf{e}^{\mathrm{del,DWC}}_{1}
&\leq
\operatorname{diag}\!\left(
\mathbf{e}^{\mathrm{traff,DWC}}_{1}
\right)
\boldsymbol{\delta}^{\mathrm{DWC}},
\qquad&
\gamma^{\mathrm{d}}_{T}\mathbf{e}^{\mathrm{del,DWC}}_{1}
&\leq
\operatorname{diag}\!\left(
\mathbf{e}^{\mathrm{cap,DWC}}
\right)
\boldsymbol{\delta}^{\mathrm{DWC}},
\label{eq:si-dwc_energy_constraints}\\
\mathbf{e}^{\mathrm{del,SCS}}_{1}
&\leq
\operatorname{diag}\!\left(
\mathbf{e}^{\mathrm{traff,SCS}}_{1}
\right)
\boldsymbol{\delta}^{\mathrm{SCS}},
\qquad&
\gamma^{\mathrm{d}}_{T}\mathbf{e}^{\mathrm{del,SCS}}_{1}
&\leq
e^{\mathrm{cap,SCS}}
\boldsymbol{\delta}^{\mathrm{SCS}},
\label{eq:si-scs_energy_constraints}
\end{alignat}

\vspace{-0.6em}
\noindent\hspace{0.165\textwidth}\textit{Variable domains:}
\vspace{-0.5em}
\begin{alignat}{3}
\boldsymbol{\delta}^{\mathrm{DWC}}
&\in\{0,1\}^{n},
\qquad&
\boldsymbol{\delta}^{\mathrm{SCS}}
&\in\{0,1\}^{m},
\qquad&
\boldsymbol{\delta}^{\mathrm{adj}}
&\in\{0,1\}^{n^{\mathrm{adj}}},
\label{eq:si-binary_domains_1}\\
\boldsymbol{\delta}^{\mathrm{sub,DWC}}
&\in\{0,1\}^{k^{\mathrm{DWC}}},
\qquad&
\boldsymbol{\delta}^{\mathrm{sub,SCS}}
&\in\{0,1\}^{k^{\mathrm{SCS}}},
\label{eq:si-binary_domains_2}\\
\mathbf{u}^{\mathrm{adj}},\mathbf{f}
&\in\mathbb{R}^{n^{\mathrm{adj}}}_{+},
\qquad&
\mathbf{u}^{\mathrm{sub}},
\mathbf{p}^{\mathrm{tap,DWC}}
&\in\mathbb{R}^{k^{\mathrm{DWC}}}_{+},
\label{eq:si-continuous_domains_1}\\
\mathbf{p}^{\mathrm{tap,SCS}}
&\in\mathbb{R}^{k^{\mathrm{SCS}}}_{+},
\qquad&
\mathbf{w}
&\in\mathbb{R}^{n^{\mathrm{sub}}}_{+},
\label{eq:si-continuous_domains_2}\\
\mathbf{e}^{\mathrm{del,DWC}}_{1}
&\in\mathbb{R}^{n}_{+},
\qquad&
\mathbf{e}^{\mathrm{del,SCS}}_{1}
&\in\mathbb{R}^{m}_{+}.
\label{eq:si-continuous_domains_3}
\end{alignat}

\end{subequations}

The objective in equation \eqref{eq:si-joint_objective} maximizes lifetime portfolio NPV by combining the present values of DWC and SCS charging revenue and subtracting electricity costs, fixed O\&M costs, and upfront capital costs. The electricity and fixed O\&M costs receive their respective lifetime growth and discounting factors, while the upfront capital cost is subtracted once at project initiation. The deployment constraint in equation \eqref{eq:si-deployment_limit} limits the combined number of newly deployed DWC segments and SCS stations to $\Pi$.

Equations \eqref{eq:si-virtual_segment_balance}--\eqref{eq:si-virtual_total_supply} impose a single-commodity virtual-flow network for DWC connectivity. Together, they require every selected DWC segment to have a valid electrical path, either directly or through adjacent deployed DWC segments, to at least one candidate substation.

Equation \eqref{eq:si-dwc_power_balance} denotes the corresponding physical power balance and equations \eqref{eq:si-dwc_substation_coupling}--\eqref{eq:si-dwc_adjacency_logic} couple the virtual and physical flows to binary infrastructure decisions. A DWC distribution line can carry flow only when it is constructed and its receiving roadway segment is selected. An adjacent link can carry virtual or physical flow only when the link is constructed and both endpoint segments are deployed as DWC. This allows a deployed segment to receive power through a neighboring electrified segment rather than always requiring its own direct substation line. Equation \eqref{eq:si-dwc_adjacency_logic} requires both endpoint roadway segments of a selected link to be deployed as DWC and prevents simultaneous selection of the two opposite directed representations of the same physical adjacent link.

Equations \eqref{eq:si-scs_connection_coupling} and \eqref{eq:si-scs_power_connection} impose the SCS electrical-connection requirements. A connection is available only for a deployed SCS site, and each deployed station is assigned its full rated station power $p^{\mathrm{SCS,stn}}$. Each candidate SCS site is connected to at most one substation.

Equations \eqref{eq:si-dwc_energy_constraints} and \eqref{eq:si-scs_energy_constraints} constrain delivered energy. Base year delivery cannot exceed the corresponding traffic-derived charging demand. The second inequality for each technology applies the terminal-year demand multiplier $\gamma^{\mathrm{d}}_{T}$ to the base year delivered-energy decision and limits the resulting year-$T$ delivery by the fixed physical hardware capacity. This reproduces the implemented treatment of lifetime demand growth while maintaining the installed charging capacity on the right-hand side.

The final set of constraints
(equations \eqref{eq:si-binary_domains_1}--\eqref{eq:si-continuous_domains_3})
specifies the domains of the optimization variables. The deployment and connection variables
$\boldsymbol{\delta}^{\mathrm{DWC}}$,
$\boldsymbol{\delta}^{\mathrm{SCS}}$,
$\boldsymbol{\delta}^{\mathrm{adj}}$,
$\boldsymbol{\delta}^{\mathrm{sub,DWC}}$, and
$\boldsymbol{\delta}^{\mathrm{sub,SCS}}$
are binary, while the virtual-flow, physical power-flow, substation-supply, and delivered-energy variables are continuous and nonnegative. The presence of both binary and continuous decision variables makes the resulting optimization problem an MILP.

\subsection*{Case study - baseline scenario}\label{sec:Framework Overview}

We use the contiguous US Interstate freight network as the case study for evaluating joint DWC and SCS deployment. We assume that all EVs are compatible with both technologies and represent charging-service allocation through exogenous traffic shares. In the baseline case, we assign equal EV traffic shares to DWC and SCS $(\eta^{\mathrm{DWC}}!=!\eta^{\mathrm{SCS}}!=!0.5)$, set the project lifetime to 30 years, and apply a 7\% annual discount rate. We assume that EV adoption grows at 15\% annually, while all retail charging prices increase by 2\% per year, CERs by 1.5\%, and fixed O\&M costs by 2\%.

We represent DWC infrastructure with an installed power density of 2~MW~mile$^{-1}$ and assign a base infrastructure cost of USD $10$ million per nominal 5 mile segment. We model each SCS as a six-port DCFC station, above the average number of ports at publicly funded US fast-charging sites \cite{NEVI2024QuarterlyUpdate}, and equip each station with six 250-kW charging ports at a base hardware cost of USD $250{,}000$ per port. We assume that both technologies operate year-round $(H^{\mathrm{op}}=8760~\mathrm{h})$. For the baseline constrained case, we set the combined deployment limit to $\Pi=200$ and omit this restriction for the unconstrained cases. We specify parameters varied across individual scenario experiments alongside the corresponding analyses and report the complete set of baseline assumptions in Supplementary Table~\ref{tab:si-baseline_parameters}.

\subsection*{Global scenario sensitivity analysis}\label{sec:Sensitivity Analysis}

We evaluate the robustness of deployment patterns to simultaneous variation in economic, technological, and demand assumptions using a global scenario sensitivity analysis. Nine parameters are varied jointly: the DWC traffic share $\eta^{\mathrm{DWC}}\in[0.10,0.90]$; the SCS traffic share $\eta^{\mathrm{SCS}}\in[0.10,0.90]$; annual traffic growth $g^{\mathrm{AADT}}\in[0.8\%,2.0\%]$; annual EV-adoption growth $g^{\mathrm{EV}}\in[10\%,18\%]$; project lifetime $T\in[25,40]$ years; DWC capital cost in USD $8$--$12$ million per segment; SCS capital cost in USD $180{,}000$--$300{,}000$ per port; 4--8 SCS ports per station; and annual CER growth in $[0.5\%,2.5\%]$. We explore the parameter space using constrained Latin-hypercube-based sampling. The DWC and SCS traffic shares are sampled separately, with infeasible combinations rejected to enforce $\eta^{\mathrm{DWC}}+\eta^{\mathrm{SCS}}\leq1$; any residual share represents EV traffic not assigned to either modeled charging technology in that scenario. The deployment-count limit is held fixed at $\Pi=200$ for all scenarios, and each accepted parameter vector is independently evaluated using the full MILP with a 1\% optimality gap.

Each scenario evaluates 9,053 candidate roadway segments and the corresponding candidate SCS sites. To isolate the factors governing technology choice, the regression sample is restricted to 198,736 deployed observations, comprising 79,006 DWC and 119,730 SCS selections. Each observation is combined with the nine scenario parameters, segment-level AADT, and nearest-substation distance, and a class-balanced, L2-regularized standardized binary logistic regression is fitted with DWC coded as 1 and SCS as 0. Positive standardized coefficients indicate conditions favoring DWC, whereas negative coefficients indicate conditions favoring SCS; undeployed segments are excluded because they do not represent a choice between the technologies. Model performance is evaluated using a grouped 80\%--20\% train--test split, with complete optimization runs assigned exclusively to either set. The model achieves a held-out balanced accuracy of 99.14\% and an area under the receiver-operating-characteristic curve of 0.9995, after which the reported coefficients are estimated using all deployed observations from the 1,000-scenario GSSA. The regression serves as an interpretable surrogate for the optimization outcomes, with its high accuracy reflecting the strong structure of the MILP decisions.

For the full analysis in Fig. \ref{fig:filteringtool}d, we fit a separate standardized binary logistic regression to selected DWC and SCS observations using segment-level AADT, CCI, CER, and nearest substation distance as predictors.


\section*{Acknowledgments}

Geospatial analyses were conducted using ArcGIS Pro 3.6.0 (Esri) under Purdue University’s institutional license. ArcGIS is the intellectual property of Esri and is used herein under license.

\section*{Funding}

This work was supported in part by the US National Science Foundation Engineering Research Center for Advancing Sustainability through Powered Infrastructure for Roadway Electrification (ASPIRE) under Award No.~1941524.

\section*{Ethics declaration}

Ethics declaration: not applicable.

\section*{Data availability}

We obtain the National Highway System (NHS) network from USDOT National Transportation Atlas Database \cite{USDOT_NTAD_NHS_2025}, with freight AADT represented using combination-truck traffic from the Highway Performance Monitoring System \cite{FHWA_HPMS_Geospatial}. Spatially resolved 2024 CERs are obtained from an Esri layer derived from US EIA Form-861 data \cite{Esri_ElectricityRates_2024,EIA_ElectricSales_2024}, a regional construction-cost proxy (CCI) is estimated from median household income and bounded between 0.7 and 1.3 \cite{EsriUpdatedDemographics2024}, and existing DCFC locations are obtained from the US Department of Energy \cite{NREL_AFDC_Stations}. We estimate DWC infrastructure and O\&M costs from Haddad et al. \cite{Haddad2022FinancialViability}, substation connection and line-extension costs from utility-reported data \cite{Eversource_DER_Project_Costs}, and SCS costs from commercial DCFC rates \cite{GreenLancer_EV_Charging_Costs_2026}. Base-year freight electrification is set to 0.5\% \cite{ICCT_RaceToZero_2025}, with 1.1\% annual freight-traffic growth \cite{FHWA_VMT_2024}. Freight charging is represented using a 300-kW DWC receiver, 250-kW SCS ports, and 150 kWh per SCS event \cite{ORNL_300kW_WPT,NLR_EVSE_Trends_2024,JointOffice_BatteryBufferedDCFC}. Both technologies use a common retail charging price of USD $0.54\;\text{kWh}^{-1}$ \cite{DCFCTracker_2026}.

\section*{Code availability}

The optimization model, sensitivity-analysis and post-processing scripts,
consolidated model inputs, validation utilities, documentation, and a
reproducible baseline example are publicly available at
\url{https://github.com/DipG02/joint-ev-charging-siting}.

\section*{Competing interests}

The authors declare no competing interests.

\section*{Author contributions}

D.G. conceived the study, developed the methodology, geospatial data workflow, and optimization model, performed all analyses, produced the figures, and wrote the manuscript. J.Q. and S.S. contributed to the conceptual framing and refinement of the methodology, supervised the research, acquired funding, provided project administration, and reviewed and edited the manuscript.


\ifdefined\COMBINEDVERSION
    \putbib[sn-bibliography]
    \clearpage
    \end{bibunit}
\else
    \bibliography{sn-bibliography}
\fi


\ifdefined\COMBINEDVERSION

    \clearpage

    \bibliographynamespace{si}
    \begin{bibunit}[sn-mathphys-num]


\hypersetup{pageanchor=false}
\begin{titlepage}
\thispagestyle{empty}

\vspace*{0.22\textheight}

\begin{center}

{\LARGE\bfseries Supplementary Information\par}

\vspace{1.2em}

{\Large\bfseries
Rethinking electrified transportation: The case for dynamic wireless charging\par
}

\vspace{1.4em}

{\large
Dipanjan Ghose$^{1}$, Junjie Qin$^{1}$, and S. Sivaranjani$^{2,*}$\par
}

\vspace{0.8em}

{\normalsize
$^{1}$Elmore Family School of Electrical and Computer Engineering,
Purdue University, West Lafayette, IN, USA.\par

$^{2}$Edwardson School of Industrial Engineering,
Purdue University, West Lafayette, IN, USA.\par

*Correspondence: sseetha@purdue.edu\par
}

\end{center}

\end{titlepage}


\setcounter{page}{1}
\renewcommand{\thepage}{\arabic{page}}
\makeatletter
\patchcmd{\Hy@EveryPageAnchor}
  {\pdfstringdef\@the@H@page{\thepage}}
  {\pdfstringdef\@the@H@page{supp.\arabic{page}}}
  {}{\PackageError{dwc}{SI page-anchor patch failed}{Check hyperref version.}}
\makeatother
\hypersetup{pageanchor=true}


\setcounter{section}{0}
\setcounter{subsection}{0}
\setcounter{subsubsection}{0}

\setcounter{figure}{0}
\setcounter{table}{0}
\setcounter{equation}{0}


\renewcommand{\thesection}{\arabic{section}}
\renewcommand{\thesubsection}{\thesection.\arabic{subsection}}
\renewcommand{\thesubsubsection}
{\thesubsection.\arabic{subsubsection}}

\renewcommand{\figurename}{Supplementary Figure}
\renewcommand{\tablename}{Supplementary Table}

\renewcommand{\thefigure}{\arabic{figure}}
\renewcommand{\thetable}{\arabic{table}}

\renewcommand{\theequation}{S\arabic{equation}}


\providecommand{\theHsection}{}
\providecommand{\theHsubsection}{}
\providecommand{\theHsubsubsection}{}
\providecommand{\theHfigure}{}
\providecommand{\theHtable}{}
\providecommand{\theHequation}{}

\renewcommand{\theHsection}
{supp.section.\arabic{section}}

\renewcommand{\theHsubsection}
{supp.section.\arabic{section}.\arabic{subsection}}

\renewcommand{\theHsubsubsection}
{supp.section.\arabic{section}.\arabic{subsection}.\arabic{subsubsection}}

\renewcommand{\theHfigure}
{supp.figure.\arabic{figure}}

\renewcommand{\theHtable}
{supp.table.\arabic{table}}

\renewcommand{\theHequation}
{supp.equation.\arabic{equation}}


\makeatletter
\newcommand{\supplementarynote}[2]{%
  \refstepcounter{section}%
  \edef\supplementarynotehref{\@currentHref}%
  \section*{Supplementary Note \thesection: #1}%
  \edef\@currentHref{\supplementarynotehref}%
  \def\@currentlabelname{#1}%
  \label{#2}%
  \addcontentsline{toc}{section}{Supplementary Note \thesection: #1}}
\renewcommand{\fnum@table}{{\bfseries\tablename\space\thetable:}}
\renewcommand{\l@section}{\@dottedtocline{1}{0em}{0em}}
\makeatother
\setcounter{tocdepth}{2}

\tableofcontents
\clearpage

\supplementarynote{GIS Workflow}{sec:si-GIS Workflow}
All GIS preprocessing and spatial analysis are performed in ArcGIS Pro 3.6.0, with all shapefile layers projected to the USA Contiguous Albers Equal Area Conic coordinate system.

\subsection{US Interstate segmentation}\label{subsec:si-US Interstate segmentation}

We use the US National Highway System (NHS) dataset \cite{USDOT_NTAD_NHS_2025} to construct the base roadway network. The original dataset contains polyline features representing Interstate highway segments. We first dissolve these roadway features using the Interstate route identifier \texttt{SIGN1} and enable the unsplit-lines option to consolidate features belonging to the same Interstate while preserving disconnected route fragments. We then repair the resulting geometries and order them geographically before segmentation. Using the ArcGIS Pro \textit{Generate Points Along Lines} tool, we place points at 5 mile intervals based on geodesic distance and include route endpoints. We subsequently apply the \textit{Split Line at Point} tool with a 10-ft search radius to divide the dissolved Interstate features at these locations, producing candidate roadway segments of approximately 5 miles. We calculate segment lengths geodesically in US miles. Supplementary Fig.~\ref{fig:si-supp_1} illustrates this segmentation process for a section of the Interstate network in central Indiana.

\begin{figure}[h]
    \centering
    \includegraphics[width=1\linewidth]{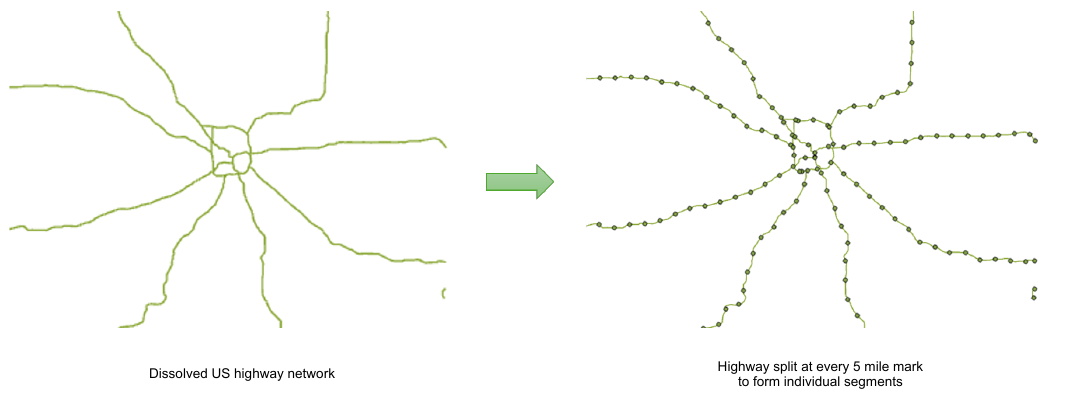}
    \caption{Illustration of the Interstate segmentation workflow in central Indiana, showing the dissolved roadway network before segmentation (left) and the resulting approximately 5 mile roadway segments defined by regularly spaced points (right).}
    \label{fig:si-supp_1}
\end{figure}

\subsection{Filtering residual roadway segments}\label{subsec:si-Filtering residual roadway segments}

The segmentation procedure produces a small number of residual roadway segments shorter than the nominal 5 mile length. These shorter segments arise primarily from overlaps and discontinuities in the original roadway geometries and from the remainder lengths created when the dissolved Interstate features are split at 5 mile intervals. To simplify the candidate network and maintain a more consistent segment scale, we remove segments shorter than 4 miles. Applying this threshold removes 1,268 segments from the initial 10,321 segmented roadway features, leaving 9,053 roadway segments for subsequent analysis. Supplementary Fig.~\ref{fig:si-supp_2} shows the removed segments across the network.

\begin{figure}
    \centering
    \includegraphics[width=0.7\linewidth]{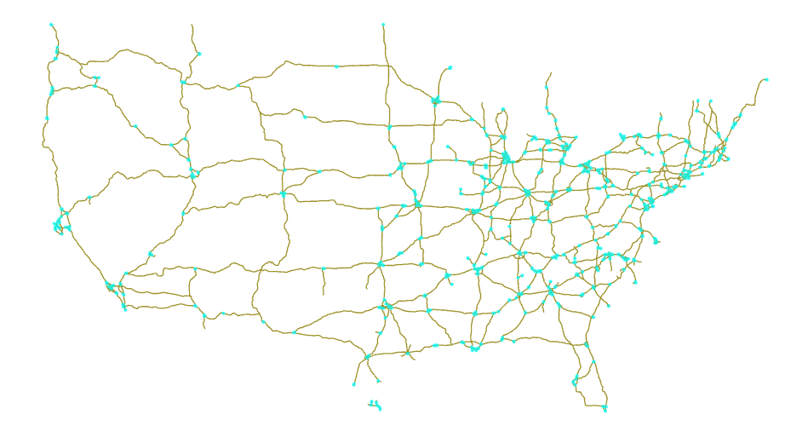}
    \caption{Roadway segments removed by the minimum-length filter. Segments shorter than 4 miles are excluded from the candidate Interstate network to remove residual fragments created by overlapping source geometries and the 5 mile segmentation procedure.}
    \label{fig:si-supp_2}
\end{figure}

\subsection{Traffic attribution to roadway segments}
\label{subsec:si-Traffic attribution to roadway segments}

We transfer traffic attributes from the original NHS roadway features to the newly generated 5 mile segments using a spatial join. The source dataset provides annual average daily traffic (AADT) fields, including total AADT, combination-truck AADT (\texttt{AADT\_COM}), and single-unit truck AADT (\texttt{AADT\_SINGL}), together with the roadway speed-limit field (\texttt{SPEED\_LIMI}). Because each newly generated segment can intersect multiple original roadway features, we perform a \texttt{JOIN\_ONE\_TO\_ONE} spatial join with the \texttt{INTERSECT} match option and use the \texttt{MAX} merge rule to retain the maximum intersecting value for each traffic-related field. This procedure preserves the highest traffic intensity represented along each candidate segment while avoiding duplicate records from overlapping source features. We replace missing traffic values after the spatial join with the median of the corresponding field across the segmented Interstate dataset. The resulting traffic-enriched roadway layer retains the unique segment identifiers assigned during segmentation and provides the segment-level traffic inputs for the subsequent charging-demand and optimization analyses. We use combination-truck AADT as the representative traffic measure for the case-study demand calculations.

\subsection{CER and CCI attribution}\label{subsec:si-CER and CCI attribution}

We next augment the segmented Interstate network with spatially resolved commercial electricity rates (CERs) and construction-cost indices (CCIs). Spatially resolved 2024 CERs are obtained from an Esri layer derived from US EIA Form-861 data \cite{Esri_ElectricityRates_2024,EIA_ElectricSales_2024} and assigned to roadway segments through spatial joins, with the total commercial electricity rate retained as the segment-level electricity-cost input. Regional CCI is represented using median household income from the Esri Updated Demographics dataset \cite{EsriUpdatedDemographics2024} as a proxy for local construction-cost variation, since a consistent national construction-cost dataset at the required spatial resolution is not available. Median household income is normalized across the study region and mapped to a bounded CCI range of 0.7--1.3. This proxy captures broad regional cost variation but does not explicitly represent terrain, permitting, or right-of-way conditions.

We replace missing or null CER and CCI values with the corresponding dataset averages. This procedure produces a comprehensive roadway dataset in which each candidate Interstate segment contains its traffic attributes, local CER, and regional CCI for use in the optimization model. Supplementary Fig.~\ref{fig:si-supp_3} illustrates the resulting Interstate network after we assign the CER and CCI attributes to the segmented roadway network.

\begin{figure}[h]
    \centering
    \includegraphics[width=0.6\linewidth]{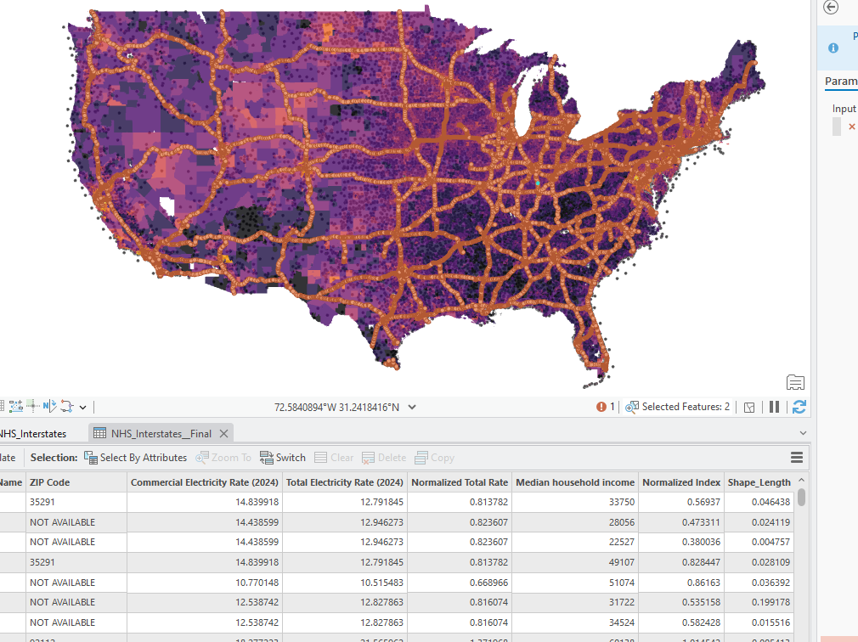}
    \caption{Comprehensive geospatial Interstate dataset after spatial attribution of commercial electricity rates and regional construction-cost information. The segmented roadway network is overlaid on the underlying spatial CER and median-household-income data used to assign segment-level CER and CCI values across the contiguous US.}
    \label{fig:si-supp_3}
\end{figure}

\subsection{Substation accessibility and grid connections}\label{subsec:si-Substation accessibility and grid connections}

We next identify candidate grid connections for each roadway segment using the geospatial substation dataset \cite{xu2020ustestsystem}. Substations located within a 5 mile radius of each candidate roadway segment are identified using the common projected coordinate system. Rather than retaining only the single closest substation, the search is performed without the ``closest feature'' restriction so that all substations within the specified radius are available as candidate grid connections. The corresponding segment--substation distances and identifiers are then retained for subsequent optimization. Supplementary Fig.~\ref{fig:si-supp_4} illustrates this filtering process for the central Indiana Interstate network, highlighting the subset of substations that fall within the specified search radius of the roadway segments.

This process produces a segment-level grid-access dataset linking each roadway segment to the set of nearby substations that can potentially supply the deployed charging infrastructure. Segments without a substation within the 5 mile search radius are retained in the roadway network but have no direct candidate grid connection. The resulting distance and connectivity information is used to construct the substation-connection inputs for the optimization model.

\begin{figure}
    \centering
    \includegraphics[width=0.32\linewidth]{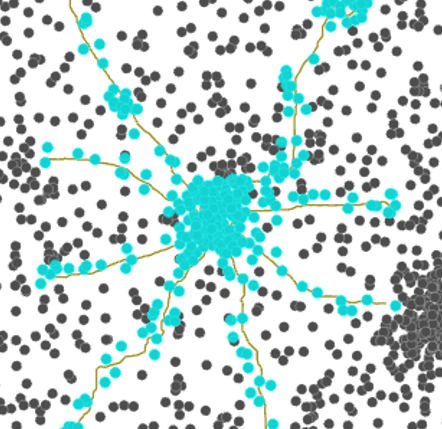}
    \caption{Identification of candidate substations for the central Indiana Interstate network. Substations located within the 5 mile search radius of the roadway segments are highlighted, while the surrounding substation population is shown for spatial context.}
    \label{fig:si-supp_4}
\end{figure}

\subsection{SCS candidate site construction}\label{subsec:si-SCS candidate site construction}

We construct candidate SCS locations from highway-ramp geometries extracted from the US OpenStreetMap roadway network \cite{OpenStreetMap}. We retain motorway-link and trunk-link features associated with the Interstate system and intersect them with the segmented Interstate network. Because the ramp geometries serve only as spatial indicators of potential stationary charging locations, we convert the relevant ramp features to point representations.

Multiple ramp points belonging to the same interchange can occur within a small area. To avoid treating these closely spaced points as separate candidate stations, we aggregate them using density-based spatial clustering (DBSCAN). Points within 400~m are grouped into a common cluster, with a minimum of two points required to form a cluster. Each resulting cluster is represented as a single candidate SCS location. Supplementary Fig.~\ref{fig:si-supp_5} illustrates this workflow for the central Indiana network, from the filtered highway network to clustered ramp points and the resulting SCS candidate sites. The resulting candidate site layer provides the spatial basis for defining the set $\mathcal{R}$ used in the optimization.

\begin{figure}[h]
    \centering
    \includegraphics[width=1\linewidth]{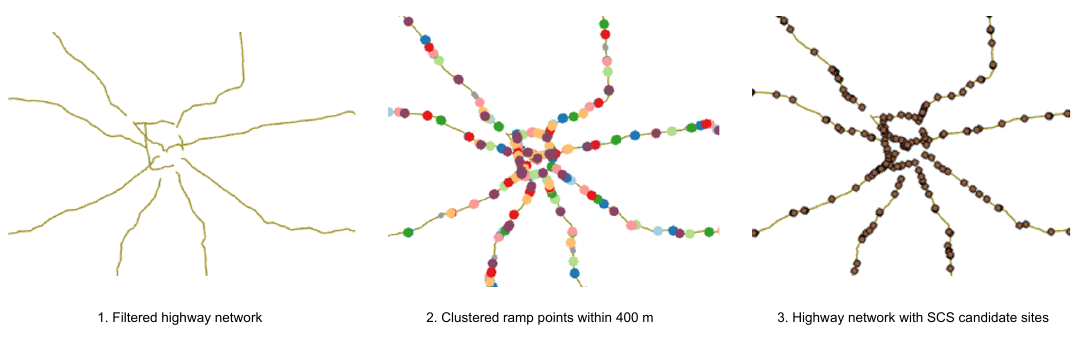}
    \caption{Construction of candidate SCS locations for the central Indiana Interstate network. The filtered highway network is first used to identify ramp locations, nearby ramp points are grouped using DBSCAN within a 400~m distance threshold, and the resulting clusters are represented as candidate SCS sites along the roadway network.}
    \label{fig:si-supp_5}
\end{figure}

\subsection{Existing DCFC infrastructure}\label{subsec:si-Existing DC fast-charging infrastructure}

We obtain existing public DCFC locations from the US Department of Energy Alternative Fuel Station dataset \cite{NREL_AFDC_Stations} and use them to identify portions of the candidate SCS network already served by charging infrastructure. We exclude candidate SCS sites located within 2 miles of an existing public DCFC station from the new-build candidate set, preventing the optimization from placing additional stationary charging infrastructure in locations with existing service.

We perform this filtering spatially by comparing the clustered SCS candidate points with existing DCFC locations and retaining only candidate sites outside the specified exclusion radius. When multiple candidate sites fall within 2 miles of the same existing charging station, we associate only the nearest candidate with that station for filtering. The resulting SCS candidate set therefore represents locations without nearby existing public DCFC access and defines the new stationary-charging deployment options considered by the optimization. We treat the SCS-bound demand associated with an excluded candidate as already served and remove it from the new-build optimization, while retaining DWC deployment eligibility on the corresponding roadway segment.

\subsection{DWC adjacency and optimization-input construction}\label{subsec:si-DWC adjacency and optimization-input construction}

We construct roadway adjacency from the segmented Interstate network to represent potential electrical connections between contiguous DWC segments. We classify two segments as adjacent only when their endpoints form a valid roadway continuation, rather than on the basis of spatial proximity alone. This criterion prevents connections between nearby but topologically disconnected roadways and preserves gaps or discontinuities in the processed network. The resulting neighboring segment pairs define the candidate DWC adjacency set used in the optimization.

We represent each physical neighboring segment pair by two directed arcs to allow power transfer in either direction and use these arcs to construct the endpoint-incidence and signed-incidence matrices defined in Methods: \nameref{sec:si-Siting MILP}. For candidate DWC connections, we retain all substations located within 5 miles of each roadway segment together with their corresponding connection distances. We map each candidate SCS site to its associated roadway segment and nearest valid candidate substation. We index the final GIS outputs consistently with the MILP sets \(\mathcal I\), \(\mathcal R\), \(\mathcal A\), \(\mathcal K^{\mathrm{DWC}}\), and \(\mathcal K^{\mathrm{SCS}}\) defined in \nameref{sec:methods}. Segment-level attributes provide the traffic, length, speed, CER, and CCI inputs, while the GIS-derived connection tables provide the assignment and incidence matrices used by the optimization. This common indexing scheme also allows us to map optimized deployment and grid-connection decisions back to the original geospatial features.

\supplementarynote{Additional Results}{sec:si-Additional Results}
\subsection{State-level variation in deployment conditions}\label{subsec:si-State-level variation in deployment conditions}

\begin{figure}[H]
    \centering
    \includegraphics[width=0.8\textwidth]{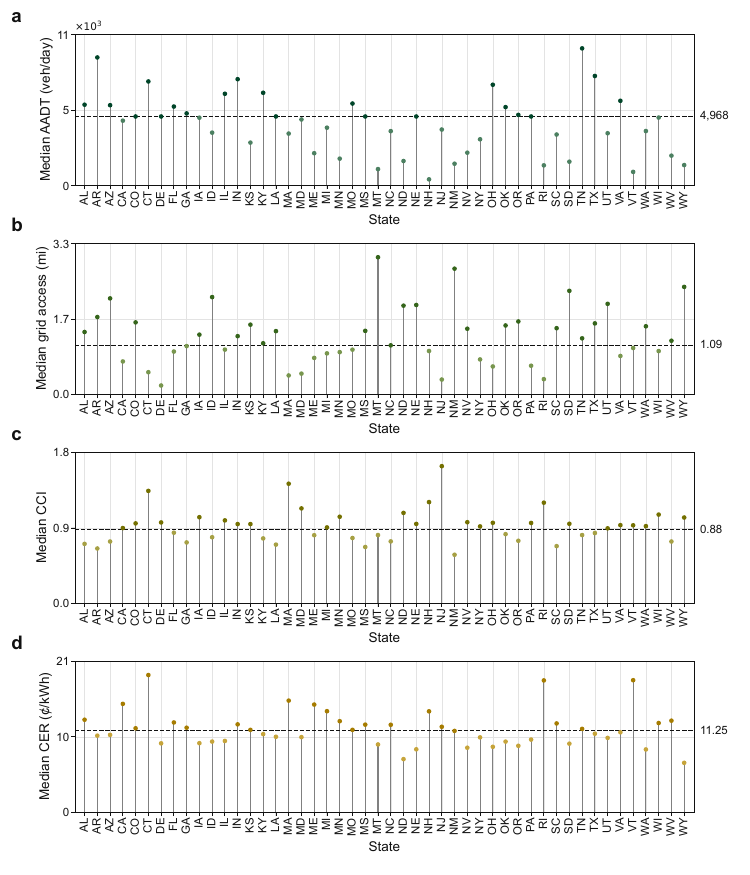}
    \caption{\textbf{State-level variation in roadway and infrastructure conditions across the US Interstate candidate network.}
Median values across candidate Interstate segments for \textbf{a}, freight AADT; \textbf{b}, grid access (nearest-substation distance); \textbf{c}, construction-cost index (CCI); and \textbf{d}, 2024 commercial electricity rate (CER). Dashed lines indicate the corresponding national medians across all candidate segments, and lighter markers denote states below the national median for each metric.}
    \label{fig:si-state_metrics}
\end{figure}

The national candidate network exhibits substantial state-level variation in the conditions that govern DWC and SCS deployment. Supplementary Fig.~\ref{fig:si-state_metrics} summarizes median freight AADT, CCI, CER, and nearest-substation distance across candidate Interstate segments in each state. The national medians are approximately 4,968 freight vehicles day$^{-1}$ for AADT, 0.88 for CCI, USD~0.1125~kWh$^{-1}$ for CER, and 1.1 miles for grid access.

Traffic and grid access show particularly strong geographic heterogeneity. States such as Tennessee, Arizona, Texas, and Illinois contain substantially higher-than-median freight traffic, whereas several western and northeastern states have considerably lower roadway utilization. Grid accessibility varies independently of traffic, with some high-demand states also benefiting from comparatively short substation distances, while others face substantially longer potential distribution line requirements for connecting to substations. Construction conditions and electricity prices introduce additional variation, but neither follows the same geographic pattern as traffic or grid access.

These differences help explain the state-level deployment outcomes reported in Figs.~3e and 5. High traffic creates the utilization needed to support DWC, while construction conditions and grid accessibility determine whether that opportunity can be realized economically. Conversely, favorable construction costs alone are insufficient to produce DWC deployment where traffic is low or grid access is poor. The state-level comparison therefore reinforces the screening interpretation in the main text: technology preference emerges from the interaction of demand, construction conditions, and grid accessibility rather than from any single regional attribute.

\subsection{Portfolio transitions underlying the breakeven surface}\label{subsec:si-Portfolio transitions underlying the breakeven surface}

The localized increases in portfolio breakeven year in Fig.~3d coincide with discrete changes in the optimized technology mix. Supplementary Fig.~\ref{fig:si-evgrowth_dwc_cost} shows the corresponding numbers of selected DWC and SCS units across the same combinations of EV-adoption growth and DWC capital-cost multiplier. At lower EV-growth rates, the fixed 200-unit portfolio remains predominantly SCS-based. As demand growth increases, DWC deployment rises rapidly and progressively displaces SCS, with higher DWC capital costs shifting this transition toward stronger EV-growth conditions.

The resulting portfolio boundary is not smooth because deployment decisions are discrete. Near the crossover, small changes in EV growth or DWC capital cost can replace a substantial number of lower-capital SCS sites with higher-capital DWC segments. These abrupt changes in portfolio composition temporarily increase the capital that must be recovered, producing the localized increases in breakeven year observed in Fig.~3d. Once DWC becomes dominant, continued demand growth improves utilization and the breakeven trajectory resumes its broader trend.

\begin{figure}[H]
    \centering
    \includegraphics[width=0.8\textwidth]{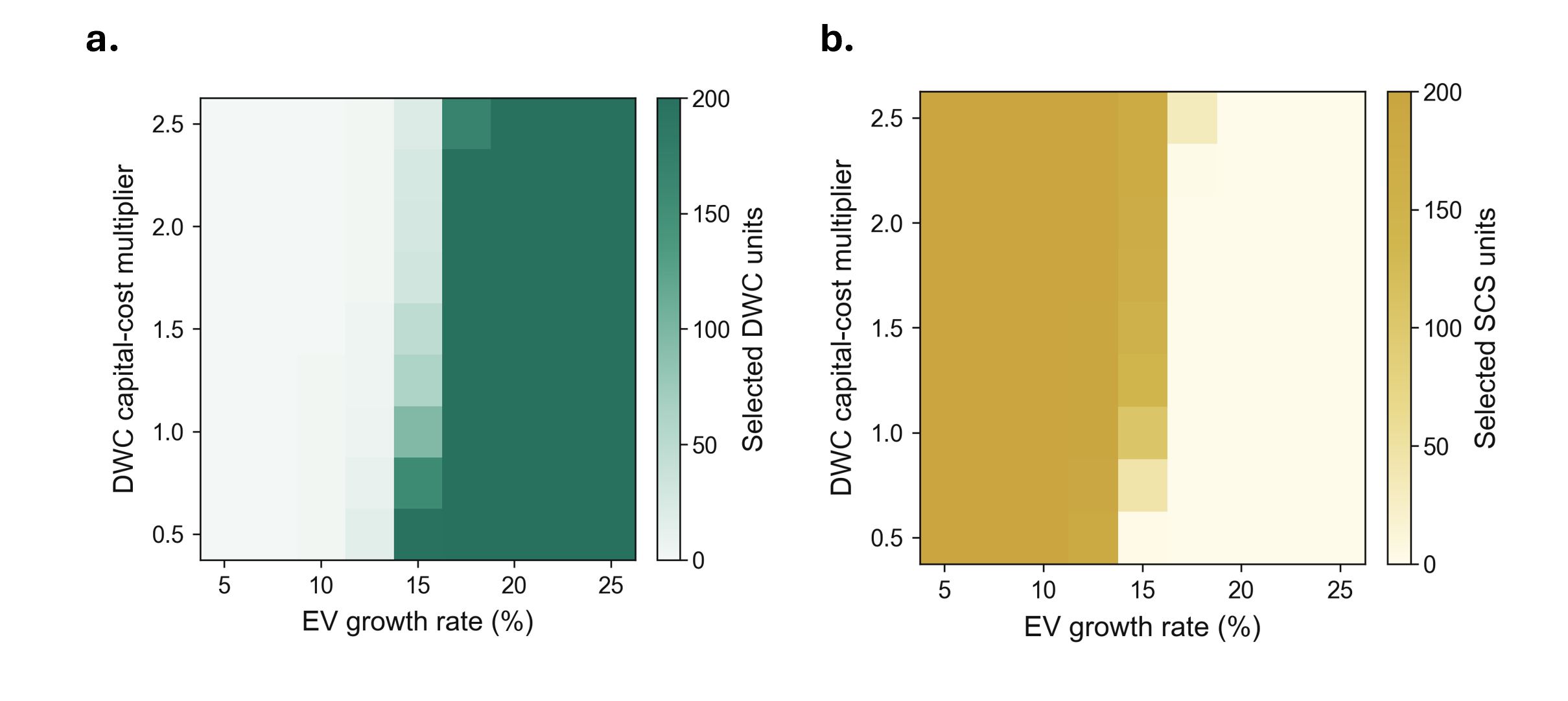}
    \caption{\textbf{Portfolio transitions underlying the breakeven-year response in Fig.~3d.}
    Numbers of selected \textbf{a}, DWC segments and \textbf{b}, SCS sites across combinations of annual EV-adoption growth and DWC capital-cost multiplier under the fixed 200-unit deployment limit. Increasing EV growth shifts the portfolio from SCS toward DWC, while higher DWC capital costs shift this transition toward stronger demand growth. The discrete change in technology composition near the transition boundary explains the localized increases in portfolio breakeven year observed in Fig.~3d.}
    \label{fig:si-evgrowth_dwc_cost}
\end{figure}

\clearpage
\section*{Supplementary Tables}
\addcontentsline{toc}{section}{Supplementary Tables}

\begin{table}[!ht]
\centering
\caption{Baseline parameters used for the national DWC--SCS case study.}
\label{tab:si-baseline_parameters}
\begin{tabular}{p{5.2cm} p{2.2cm} p{2.8cm}}
\hline
\textbf{Parameter} & \textbf{Baseline value} & \textbf{Unit} \\
\hline

\multicolumn{3}{l}{\textit{Demand and deployment}} \\[2pt]
Initial EV share of freight AADT, $\rho^{\mathrm{EV}}_1$ & 0.5 & \% \\
DWC traffic share, $\eta^{\mathrm{DWC}}$ & 0.50 & -- \\
SCS traffic share, $\eta^{\mathrm{SCS}}$ & 0.50 & -- \\
Annual AADT growth, $g^{\mathrm{AADT}}$ & 1.1 & \% yr$^{-1}$ \\
Annual EV-adoption growth, $g^{\mathrm{EV}}$ & 15 & \% yr$^{-1}$ \\
Combined deployment-count limit, $\Pi$ & 200 & deployments \\
Project lifetime, $T$ & 30 & yr \\

\multicolumn{3}{l}{\textit{Economic parameters}} \\[2pt]
Discount rate, $\rho$ & 7 & \% yr$^{-1}$ \\
DWC charging price, $c^{\mathrm{rev,DWC}}$ & 0.54 & USD kWh$^{-1}$ \\
SCS charging price, $c^{\mathrm{rev,SCS}}$ & 0.54 & USD kWh$^{-1}$ \\
Annual DWC charging price growth, $g^{\mathrm{p,DWC}}$ & 2.0 & \% yr$^{-1}$ \\
Annual SCS charging price growth, $g^{\mathrm{p,SCS}}$ & 2.0 & \% yr$^{-1}$ \\
Annual CER growth, $g^{\mathrm{c,CER}}$ & 1.5 & \% yr$^{-1}$ \\
Annual fixed O\&M growth, $g^{\mathrm{c,om}}$ & 2.0 & \% yr$^{-1}$ \\
CCI bounds & 0.70--1.30 & -- \\

\multicolumn{3}{l}{\textit{DWC technology and cost}} \\[2pt]
Nominal DWC segment length & 5 & miles \\
DWC infrastructure cost per nominal segment, $c^{\mathrm{DWC}}$ & 10.0 & USD million \\
Internal electrical-line cost, $c^{\mathrm{int}}$ & 70,000 & USD segment$^{-1}$ \\
Adjacent-link cost, $c^{\mathrm{adj}}$ & 400,000 & USD link$^{-1}$ \\
Distribution line cost (for substation connections), $c^{\mathrm{ext}}$ & 1.39 & USD million mile$^{-1}$ \\
Grid-connection cost, $c^{\mathrm{conn}}$ & 0.60 & USD million MW$^{-1}$ \\
Installed DWC power density, $p^{\mathrm{coil}}$ & 2.0 & MW mile$^{-1}$ \\
EV-side DWC receiver power, $p^{\mathrm{DWC}}$ & 300 & kW \\

\multicolumn{3}{l}{\textit{SCS technology and cost}} \\[2pt]
Charging ports per SCS, $n^{\mathrm{SCS}}$ & 6 & ports station$^{-1}$ \\
Rated power per SCS port, $p^{\mathrm{SCS}}$ & 250 & kW \\
Installed SCS station power & 1.5 & MW station$^{-1}$ \\
SCS capital cost per port, $c^{\mathrm{port}}$ & 250,000 & USD port$^{-1}$ \\
Base SCS hardware cost & 1.50 & USD million station$^{-1}$ \\
Energy per captured SCS charging event, $e^{\mathrm{req}}$ & 150 & kWh event$^{-1}$ \\
SCS utilization factor, $\mu^{\mathrm{SCS}}$ & 1.0 & -- \\
Base fixed O\&M fraction & 1.5 & \% of installed infrastructure and grid-connection cost yr$^{-1}$ \\
Annual operating duration, $H^{\mathrm{op}}$ & 8,760 & h yr$^{-1}$ \\

\hline
\end{tabular}
\end{table}
\clearpage

    \renewcommand{\refname}{Supplementary references}
    \putbib[sn-bibliography]
    \clearpage
    \end{bibunit}

\fi

\end{document}